\documentclass[useAMS,usenatbib]{mn2e}
\usepackage{epsfig}
\usepackage{graphics}
\usepackage[usenames]{color}
\usepackage[T1]{fontenc}

\usepackage{amssymb}
\usepackage{amsmath}
\usepackage{times}
\newcommand{\be}{\begin{equation}}
\newcommand{\ee}{\end{equation}}
\newcommand{\bdm}{\begin{displaymath}}
\newcommand{\edm}{\end{displaymath}}
\newcommand{\bea}{\begin{eqnarray}}
\newcommand{\eea}{\end{eqnarray}}

\newcommand{\cR}[1]{\textcolor{black}{#1} }

\newcommand{\msun}{M_\odot}
\def\lsim{\lower.5ex\hbox{$\; \buildrel < \over \sim \;$}}
\def\mF{\mathcal{F}}

\def\be{\begin{equation}}
\def\en{\end{equation}}
\def\bea{\begin{eqnarray}}
\def\ena{\end{eqnarray}}

\title[EPTA limits]
{European Pulsar Timing Array Limits on Continuous Gravitational Waves from Individual Supermassive Black Hole Binaries}

\author[S.~Babak  et al.]{\parbox{\textwidth}{S. Babak$^{1}$\thanks{E-mail: Stanislav.babak@aei.mpg.de}, 
       A.~Petiteau$^{2}$, 
       A.~Sesana$^{3,1}$,
       P.~Brem$^{1}$,
       P.~A.~Rosado$^{4,5}$,
       S.~R.~Taylor$^{6,7}$,
       A.~Lassus$^{8,9}$,  
       J.~W.~T. Hessels$^{10,13}$,     
        C.~G.~Bassa$^{10, 14}$,
        M.~Burgay$^{11}$,
        R.~N.~Caballero$^{8}$,
        D.~J.~Champion$^{8}$,
         I.~Cognard$^{9,12}$,
         G.~Desvignes$^{8}$,
         J.~R.~Gair$^{7}$,   
       L.~Guillemot$^{9,12}$,
       G.~H.~Janssen$^{10, 14}$,
       R.~Karuppusamy$^{8}$,
       M.~Kramer$^{8,14}$
       P.~Lazarus$^{8}$,
       K.J.~Lee$^{15}$,
       L.~Lentati$^{16}$,       
       K.~Liu$^{8}$,
        C.~M.~F.~Mingarelli$^{17, 8, 3}$,
       S.~Os{\l}owski$^{18,8}$,
       D.~Perrodin$^{11}$,
       A.~Possenti$^{11}$,
       M.~B.~Purver$^{14}$,
       S.~Sanidas$^{13,14}$,
       R.~Smits$^{10}$,
       B.~Stappers$^{14}$,
       G.~Theureau$^{9,12,19}$,
       C.~Tiburzi$^{20,11}$,
       R.~van~Haasteren$^{17}$,
       A.~Vecchio$^{3}$,        
       J.~P.~W.~Verbiest$^{18,8}$ }\vspace{0.4cm}\\ %
$^{1}$ Max-Planck-Institut f{\"u}r Gravitationsphysik, Albert-Einstein-Institut, Am M\"uhlenberg 1, 14476, Golm, Germany \\
$^{2}$ Universit\'e Paris-Diderot-Paris7 APC - UFR de Physique, B\^atiment Condorcet ,10 rue Alice Domont et L\'eonie Duquet 75205 PARIS CEDEX 13, France\\
$^{3}$ School of Physics and Astronomy, University of Birmingham, Edgbaston, Birmingham, B15 2TT, United Kingdom \\
$^{4}$ Centre for Astrophysics \& Supercomputing, Swinburne University of Technology, PO Box 218, Hawthorn VIC 3122, Australia \\
$^{5}$ Max-Planck-Institut f\"ur Gravitationsphysik, Albert-Einstein-Institut, Callinstra\ss e 38, 30167, Hannover, Germany \\
$^{6}$ Jet Propulsion Laboratory, California Institute of Technology, Pasadena, California 91109, USA\\
$^{7}$ Institute of Astronomy, University of Cambridge, Madingley Road, Cambridge, CB3 0HA, UK \\
$^{8}$ Max-Planck-Institut f{\"u}r Radioastronomie, Auf dem H{\"u}gel 69, D-53121 Bonn, Germany\\
$^{9}$ Laboratoire de Physique et Chimie de l'Environnement et de l'Espace LPC2E CNRS-Universit{\'e} d'Orl{\'e}ans, F-45071 Orl{\'e}ans, France \\
$^{10}$ ASTRON, the Netherlands Institute for Radio Astronomy, Postbus 2, 7990 AA, Dwingeloo, The Netherlands \\
$^{11}$ INAF - Osservatorio Astronomico di Cagliari, via della Scienza 5, I-09047 Selargius (CA), Italy \\
$^{12}$ Station de radioastronomie de Nan{\c c}ay, Observatoire de Paris, CNRS/INSU F-18330 Nan{\c c}ay, France \\
$^{13}$ Anton Pannekoek Institute for Astronomy, University of Amsterdam, Science Park 904, 1098 XH Amsterdam, The Netherlands \\
$^{14}$ Jodrell Bank Centre for Astrophysics, University of Manchester, Manchester, M13 9PL, United Kingdom\\
$^{15}$ Kavli institute for astronomy and astrophysics,Peking University, Beijing 100871,P.R.China\\
$^{16}$ Astrophysics Group, Cavendish Laboratory, JJ Thomson Avenue, Cambridge, CB3 0HE, UK \\
$^{17}$ TAPIR (Theoretical Astrophysics), California Institute of Technology, Pasadena, California 91125, USA \\
$^{18}$ Fakult\"at f\"ur Physik, Universit\"at Bielefeld, Postfach 100131, 33501 Bielefeld, Germany \\
$^{19}$ Laboratoire Univers et Th\'eories LUTh, Observatoire de Paris, CNRS/INSU, Universit\'e Paris Diderot, 5 place Jules Janssen, 92190 Meudon, France \\
  $^{20}$ Dipartimento di Fisica - Universit\'a di Cagliari, Cittadella Universitaria, I-09042 Monserrato (CA), Italy \\
}

\begin{document}

\date{}

\pagerange{\pageref{firstpage}--\pageref{lastpage}} \pubyear{2010}

\maketitle

\label{firstpage}

\begin{abstract}

We have searched for continuous gravitational wave (CGW) signals
produced by individually resolvable, circular supermassive black hole
binaries (SMBHBs) in the latest EPTA dataset, which consists of
ultra-precise timing data on 41 millisecond pulsars. We develop
frequentist and Bayesian detection algorithms to search both for
monochromatic and frequency-evolving systems. None of the adopted
algorithms show evidence for the presence of such a CGW signal,
indicating that the data are best described by pulsar and radiometer noise
only. Depending on the adopted detection algorithm, the 95\% upper
limit on the sky-averaged strain amplitude lies in the range $6\times
10^{-15}<A<1.5\times10^{-14}$ at $5{\rm nHz}<f<7{\rm nHz}$. 
This limit varies by a factor of five,
  depending on the assumed source position, and the most constraining
  limit is achieved towards the positions of the most sensitive
  pulsars in the timing array.
The most robust upper limit -- obtained via a full Bayesian analysis searching
simultaneously over the signal and pulsar noise on the subset of ours
six best pulsars -- is $A\approx10^{-14}$. 
These limits, the most stringent to date at $f<10{\rm nHz}$,
exclude the presence of sub-centiparsec binaries
with chirp mass $\mathcal{M}_c>10^9\msun$ out to a distance of about
25Mpc, and with $\mathcal{M}_c>10^{10}\msun$ out to a distance of
about 1Gpc ($z\approx0.2$). We show that state-of-the-art SMBHB
population models predict $<1\%$ probability of detecting a CGW with
the current EPTA dataset, consistent with the reported non-detection.
We stress, however, that PTA limits on individual CGW have improved by
almost an order of magnitude in the last five years. The continuing
advances in pulsar timing data acquisition and analysis techniques
will allow for strong astrophysical constraints on the population of nearby
SMBHBs in the coming years.
\end{abstract}


\newpage

\section{Introduction}
\label{sec:Introduction}

The direct detection of gravitational waves (GWs) is one of the
primary goals of contemporary observational astrophysics. The access
to GW information alongside  well established electromagnetic
observations will be a milestone in our investigation of the Universe,
opening the era of multimessenger astronomy.

Precision timing of an array of millisecond pulsars (i.e. a pulsar
timing array, PTA) provides a unique opportunity to get the very first
low-frequency (nHz) GW detection. PTAs exploit the effect of GWs on
the propagation of radio signals from ultra-stable millisecond pulsars
(MSPs) to the Earth
\citep[e.g.][]{1978SvA....22...36S,1979ApJ...234.1100D}, producing a
characteristic fingerprint in the times of arrival (TOAs) of radio
pulses. In the timing analysis, TOAs are fitted to a physical model
accounting for all the known processes affecting the generation,
propagation and detection of the radio pulses. { The timing residuals
are the difference between the observed TOAs and the 
TOAs predicted by the best-fit model, and they carry information about unaccounted noise
and potentially unmodelled physical effects, such as 
GWs, in the datastream}
\citep[e.g.][]{1983ApJ...265L..39H,2005ApJ...625L.123J}. The European
Pulsar Timing Array \cite[EPTA,~][]{2013CQGra..30v4009K}, the Parkes
Pulsar Timing Array \cite[PPTA,~][]{2013CQGra..30v4007H} and the North
American Nanohertz Observatory for Gravitational Waves
\cite[NANOGrav,~][]{2013CQGra..30v4008M}, joining together in the
International Pulsar Timing Array
\cite[IPTA,~][]{2010CQGra..27h4013H,2013CQGra..30v4010M}, are
constantly improving their sensitivity in the frequency range of
$\sim10^{-9}-10^{-6}$ Hz.

The primary GW source in the nHz window is a large population of
adiabatically inspiralling supermassive black hole binaries (SMBHBs),
formed following the frequent galaxy mergers occurring in the Universe
\citep{1980Natur.287..307B}. Signals from a cosmic string network
\citep[see, e.g.][]{1981PhLB..107...47V,1994csot.book.....V} or from
other physical processes occurring in the early Universe \citep[see,
  e.g.][]{2005PhyU...48.1235G} are also possible, but we will
concentrate on SMBHBs in this paper. Consisting of a superposition of
several thousands of sources randomly distributed over the sky
\citep{2008MNRAS.390..192S}, the signal has classically been described
as a stochastic GW background
\citep[GWB,][]{1995ApJ...446..543R,2003ApJ...583..616J,2003ApJ...590..691W,2004ApJ...611..623S}. Consequently,
in the last decade several detection techniques have been developed in
this direction
\citep[e.g.][]{2009PhRvD..79h4030A,2009MNRAS.395.1005V,2013PhRvD..87j4021L,2014arXiv1410.8256C}
and applied to the EPTA, PPTA, and NANOGrav datasets to get limits on
the amplitude of a putative isotropic GWB
\citep{2006ApJ...653.1571J,2011MNRAS.414.1777Y,2011MNRAS.414.3117V,2013ApJ...762...94D,2013Sci...342..334S,EPTA4isotropic}.

However, \cite{2009MNRAS.394.2255S} \citep[see
  also][]{2012ApJ...761...84R} first showed that the signal is
dominated by a handful of sources, some of which might be individually
resolvable. The typical evolution timescale of those SMBHBs is
thousands-to-millions of years, far exceeding the observational
baseline of PTA experiments (about two decades); therefore, their
signals can be modeled as non-evolving continuous GWs
\citep[CGWs,][]{2010PhRvD..81j4008S}. Resolvable sources are
particularly appealing because, if detected and localized on the sky,
they can also be followed up electromagnetically, thus providing a
multimessenger view
\citep{2012MNRAS.420..860S,2012MNRAS.420..705T,2013CQGra..30v4013B,2014MNRAS.439.3986R}.

This prospect triggered a burst of activity in the development of
search and parameter estimation algorithms for CGWs from circular SMBHBs
\citep{2010PhRvD..81j4008S,2011MNRAS.414.3251L,2012PhRvD..85d4034B,2012ApJ...753...96E,2013PhRvD..87f4036P,taylorellisgair14}, and more recently led to the development of the  first pipelines for eccentric binaries \citep{thgm15,zhu2015b}. The
pioneering work of \cite{2010MNRAS.407..669Y} was the first to produce
sensitivity curves and set upper limits using a power spectral
summation method. More recently, \cite{2014ApJ...794..141A} applied
\cR{the frequentist and Bayesian methods for evolving and non-evolving signals described} in \cite{2012ApJ...753...96E} to the NANOGrav
5-year dataset \citep{2013ApJ...762...94D}, whereas
\cite{2014MNRAS.444.3709Z} applied a frequentist method to the PPTA
data release (DR1) presented in \cite{2013PASA...30...17M}. Those
limits are usually cast in terms of the intrinsic strain amplitude of
the wave, $h_0$ or its inclination-averaged version (which is a
factor of 1.26 larger)
as a function of frequency, both averaged over the entire sky or as a function of sky location. The best sky-averaged 95\% confidence upper limit on $h_0$ quoted to date is $1.7\times10^{-14}$ at 10nHz \citep{2014MNRAS.444.3709Z}.  

Here we investigate the presence of non-evolving continuous waves from circular binaries in
the latest EPTA data release \citep{EPTA1data}. We perform a
comprehensive study applying both frequentist
\citep{2012PhRvD..85d4034B,2012ApJ...753...96E,2013PhRvD..87f4036P}
and Bayesian \citep{Ellis:2013hna,taylorellisgair14,Lassus15} methods,
and searching for both evolving and non-evolving GW signals. The paper is
organized as follows. In Section \ref{sec:GWmodel} we introduce the
EPTA dataset and the adopted gravitational waveform model. Section
\ref{sec:methods} is devoted to the description of the techniques
developed to analyze the data, divided into frequentist and Bayesian
methods. Our main results (upper limit, sensitivity curves, sky maps)
are presented in Section \ref{sec:results}, and their astrophysical
interpretation is discussed in Section \ref{sec:astro}. Finally, we
summarize our study in Section \ref{sec:conclusions}. Throughout the
paper we use geometrical units $G=c=1$.

This research is the result of the common effort to directly detect
gravitational waves using pulsar timing, known as the European Pulsar
Timing Array
\citep{2013CQGra..30v4009K}\footnote{www.epta.eu.org/}.

\section{EPTA dataset and gravitational wave model}
\label{sec:GWmodel}

\subsection {The EPTA Dataset}
\label{subsec:dataset}

In this paper we make use of the full EPTA data release described in
\cite{EPTA1data}, which consists of 42 MSPs monitored
for timespans ranging from 7 to 24 years. However, we exclude PSR
J1939+2134 from our analysis because it shows a large, unmodelled
red-noise component in its timing residuals. The remaining 41 MSPs
show well-behaved rms residuals between 130ns and 35$\mu$s. For each of
these pulsars, a full timing analysis has been performed using a
time-domain Bayesian method based on MultiNest \citep{Feroz:2008xx},
which simultaneously includes the white noise modifiers EFAC and EQUAD\footnote{
 EFAC is used  to account for possible mis-calibration of the radiometer noise and it acts as a multiplier for all the 
TOA error bars. EQUAD represents
some additional (unaccounted) source of time independent noise and it is added in 
quadratures to the TOA error bars.}
for each observing system, as well as intrinsic red noise and (observational)
frequency dependent dispersion measure (DM)
variations. Variations in the DM are due to a changing line-of-sight
  through the interstellar medium towards the pulsar. Hereinafter, we refer to this timing analysis as single pulsar analysis (SPA). As a sanity check, parallel analyses have also
been done in the frequency domain using the TempoNest plugin
\citep{2014MNRAS.437.3004L} for the Tempo2 pulsar timing package
\citep{2006MNRAS.369..655H}, and in time domain using  search
method combining a genetic algorithm~\cite[]{Petiteau:2010zu} with MCMCHammer \cite[]{mcmchammer}. The results of the three methodologies
are consistent.

For the searches performed in this paper,
we use the results of the SPA produced by MultiNest. Those consist of
posterior probability distributions for the relevant noise parameters
(EFAC, EQUAD, DM and intrinsic red noise), together with their maximum
likelihood (ML) values. \cR{Extensive noise analyses on the same dataset
are fully detailed in \cite{EPTA2DM, EPTA3Noises}, and the posterior
distributions of the noise parameters of the three most sensitive
pulsars in our array (J1909$-$3744, J1713$+$0747, J1744$-$1134) are also given in 
Figure 3 of the companion paper \cite{EPTA4isotropic}.}
In most cases we fix the noise parameters to
their ML value, but we also perform separate searches sampling from
the posterior distribution or searching simultaneously over the signal
and the noise parameters, as described in Section \ref{sec:methods}
and summarized in Table \ref{tab2}.


\subsection{Gravitational wave model}
\label{subsec:GWmodel}
{  In this subsection we introduce the mathematical description of the GW signal from a binary in circular orbit and the associated PTA response. The main purpose here is to introduce the notation; we refer the reader to \cite{2010PhRvD..81j4008S} for a full derivation of the relevant equations}.
The timing residuals of radio pulses due to the propagation of the electromagnetic waves in the field of an intervening GW can be written as
\be
r_a(t) = \int_0^{t} \frac{\delta \nu_a}{\nu_a}(t') dt',
\ee
where
\be
\frac{\delta \nu_a}{\nu_a} = \frac{1}{2} \frac{\hat{p}_a^i \hat{p}_a^j}{1 + \hat{p}_a.\hat{\Omega}} \Delta h_{ij}.
\label{dnuovernu}
\ee
{Here $\nu_a$ is the frequency of the TOAs \cR{(i.e., the spin frequency of the pulsar)} and $\delta \nu_a$ its deviation.}
The index $a$ labels a particular pulsar ($a=1,...,N$ where $N=41$ is the number of the pulsars in our array) and indicates that a given quantity explicitly depends on the parameters of the individual pulsar, $\hat{p}$ denotes the position of the pulsar on the sky, and $\hat{\Omega}$ is the direction of the GW propagation. The last factor in equation (\ref{dnuovernu}) depends on the strain of the GW at the location of the pulsar $h_{ij}(t_a^p)$ and on Earth $h_{ij}(t)$:
\be
\Delta h_{ij} = h_{ij}(t_a^p) - h_{ij}(t).
\en
The pulsar time $t_a^p$ is related to the Earth time $t$ as:
\be
t_a^p = t - L_{a}(1 + \hat{\Omega}.\hat{p}_{a}) \equiv t - \tau_{a},
\en
where $L_{a}$ is the distance to the pulsar. We consider a non-spinning binary system in quasi-circular orbit. To leading order, the response of a particular pulsar to a passing GW (that is, the induced timing residual) is given by: 
\be
r_{a}(t) = r_{a}^p(t)-r_{a}^e(t), 
\en
where
\bea
r_{a}^e(t) = \frac{\mathcal{A}}{\omega} \left\{
(1 + \cos^2{\iota}) F_{a}^{+} \left[ \sin(\omega t + \Phi_0)  - \sin{\Phi_0}\right] + \right. \nonumber \\
\left. 2 \cos{\iota} F_{a}^{\times} \left[ \cos(\omega t + \Phi_0)  - \cos{\Phi_0}\right]
\right\}, \nonumber \\
r_{a}^p(t) = \frac{\mathcal{A}_{a}}{\omega_{a}} \left\{
(1 + \cos^2{\iota}) F_{a}^{+} \left[ \sin(\omega_{a} t + \Phi_{a} + \Phi_0)  - \right.\right. \nonumber \\
\left.\left.  \sin({\Phi_{a} +\Phi_0})\right] +
 2 \cos{\iota} F_{a}^{\times} \left[ \cos(\omega_{a}t + \Phi_{a} + \Phi_0)  -     \right. \right. \nonumber \\
\left. \left.  \cos(\Phi_{a} + \Phi_0)\right] \right\}. \nonumber \\
\label{Eq:signal}
\ena
Equation (\ref{Eq:signal}) contains all the relevant features of the signal, and deserves some detailed explanation.  Here $\iota$ is the inclination of the SMBHB orbital plane with respect to the line-of-sight, and $\mathcal{A}$ (sometimes referred to as $h_0$ in the PTA literature) is the amplitude of the GW strain given by
\be
\mathcal{A} = 2\frac{\mathcal{M}_c^{5/3}}{D_L} (\pi f)^{2/3}.
\label{Eq:amplitude}
\en
Throughout the paper we consider a SMBHB with {\it redshifted} chirp mass $\mathcal{M}_c = M\eta^{3/5}$, where $M=(m_1+ m_2)$ and $\eta=m_1 m_2/M^2$ are the total mass and symmetric mass ratio of the binary system; $D_L$ is the luminosity distance to the source, and $f=\omega/(2\pi)$ is the {\it observed} GW frequency, which is twice the SMBHB observed orbital frequency{\footnote{The relation between redshifted chirp mass and rest-frame chirp mass is $\mathcal{M}_c =(1+z)\mathcal{M}_{c,{\rm rf}}$; likewise, the relation between observed and rest-frame frequency is $f_{\rm rf}=(1+z)f$. The GW community works with redshifted quantities because this is what is directly measured in the observations, and because this way, the $1+z$ factors cancel out in the equations, making the math cleaner. We note, moreover, that current PTAs might be able to resolve SMBHBs out to $z\sim0.1$, implying only a small difference between rest-frame and redshifted quantities.}}. We specify $\mathcal{A}$ and $\mathcal{A}_{a}$ in equation (\ref{Eq:signal}) because the GW frequency might be different in the pulsar and Earth terms, implying different amplitudes. In fact, in the quadrupole approximation, the evolution of the binary orbital frequency $\omega_{\rm orb}=\omega/2$ and GW phase can be written as
\be
\omega_{\rm orb}(t)=\omega_{\rm orb}\left(1-\frac{256}{5}{\mathcal{M}}^{5/3}\omega_{\rm orb}^{8/3}t\right)^{-3/8},
\label{Eq:freqev}
\en

\be
\Phi(t)=\Phi_0+\frac{1}{16{\mathcal{M}}^{5/3}}\left(\omega_{\rm orb}^{-5/3}-\omega_{\rm orb}(t)^{-5/3}\right).
\label{Eq:phaseev}
\en
In equation (\ref{Eq:signal}), $\omega$ and $\omega_{a} =
\omega(t-\tau_{a})$ are the GW frequencies of the Earth term and
pulsar term, respectively. Over the typical duration of a PTA
experiment (decades), these two frequencies can be approximated as
constants \citep[see, e.g.][]{2010PhRvD..81j4008S}, and we drop the
time dependence accordingly. However, the delay $\tau_{a}$ between the
pulsar and the Earth term is of the order of the pulsar-Earth light
travel time, and can be thousands of years. This is comparable with
the evolutionary timescale of the SMBHB's orbital frequency
\citep{2010PhRvD..81j4008S}. In particular $\omega_{a}=2\omega_{\rm
  orb,a}$ is obtained by setting $t=-\tau_{a}$ in the right-hand side
of equation (\ref{Eq:freqev}), and can be generally different from one
pulsar to the other and from $\omega$. Combining equations
(\ref{Eq:freqev}) and (\ref{Eq:phaseev}), one can show that $\Phi_a
\approx -\omega\tau_a$.

If $T$ is the timing baseline of a given PTA's set of observations, then its nominal Fourier frequency resolution bin is given by $\Delta{f}\approx1/T$. We therefore have two possibilities:
\begin{itemize}
\item If $(\omega_a-\omega)/(2\pi)>\Delta{f}$ for the majority of the MSPs in the array, then the pulsar and the Earth terms fall in different frequency bins; all the Earth terms can be added-up coherently and the pulsar terms can be considered either as separate components of signal or as an extra incoherent source of noise.
\item Conversely, if $(\omega_a-\omega)/(2\pi)<\Delta{f}$ for the majority of the MSPs, then the pulsar terms add up to the respective Earth terms, destroying their phase coherency.  
\end{itemize}
This distinction has an impact on the detection methodology that should be adopted, as we will see in the next Section.

 The antenna response functions  $F^{+,\times}_{a}$  of each  pulsar to the GW signal depend on the mutual pulsar-source position in the sky and are given by:
\bea
F_{a}^{+} &=&  \frac{1}{2} \frac{(\hat{p}^{a}. \hat{u})^2 -  (\hat{p}^{a}. \hat{v})^2}{1 + \hat{p}^{a}. \hat{\Omega}},\\
F_{a}^{\times} &=& \frac{(\hat{p}^{a}. \hat{u}) (\hat{p}^{a}. \hat{v})}{1 + \hat{p}^{a}. \hat{\Omega}},
\label{Eq:pattern}
\ena
where 
\bea
\hat{\Omega} &=& -\{\sin{\theta_S} \cos{\phi_S}, \sin{\theta_S} \sin{\phi_S}, \cos{\theta_S} \},\\
\hat{u} &=& \hat{n}\cos{\psi}  - \hat{m}\sin{\psi}, \;\;\; \hat{v} = \hat{m}\cos{\psi}  + \hat{n}\sin{\psi},\\
\hat{n} &=& \{\cos{\theta_S} \cos{\phi_S}, \cos{\theta_S} \sin{\phi_S}, -\sin{\theta_S} \},\\
\hat{m} &=& \{\sin{\phi_S}, -\cos{\phi_S}, 0 \}.
\ena
Here $\theta_S,\phi_S$ define the source sky location (respectively latitude and longitude), and $\psi$ is the GW polarization angle.

\subsection{Likelihood function and noise model}
\label{subsec:Likelihood}
The core aspect of all searches performed in this study (both frequentist and Bayesian) is the evaluation of the likelihood that some signal described by equation (\ref{Eq:signal}) is present in the time series of the pulsar TOAs. We use the expression for the likelihood marginalised over the timing parameters as described in detail in \cite{2013MNRAS.428.1147V}:
\begin{eqnarray}
        \mathcal{L}(\vec{\theta}, \vec{\lambda} | \vec{\delta t}) = \frac{1}{\sqrt{(2\pi)^{n-m} det(G^T C G) }} \times
\nonumber \\
\exp{\left(-\frac1{2} (\vec{\delta t} - \vec{r})^T G (G^T C G)^{-1} G^T (\vec{\delta t} - \vec{r}) \right)}.
\label{Eq:lik}
\end{eqnarray}
Here  $n$ is the length of the vector $\vec{\delta t} = \cup x_{a}$ obtained by concatenating the individual pulsar TOA series $x_{a}$, $m$ is the total number of parameters describing the timing model, and the matrix $G$ is related to the design matrix \citep[see][for details]{2013MNRAS.428.1147V}. The variance-covariance matrix $C$, in its more general version, contains contributions from the GWB and from the white and (in general) red noise: $C = C_{gw} + C_{wn} + C_{rn}$. We refer the reader to \cite{2009MNRAS.395.1005V} and to our companion paper \cite{EPTA4isotropic} for exact expressions of the noise variance-covariance matrix. In our analysis we use both the time \citep{2013MNRAS.428.1147V} and frequency domain \citep{2013PhRvD..87j4021L} representation of this matrix. Both approaches give qualitatively and quantitatively  consistent results (as we checked during our analysis).  Therefore, we will not specify which representation was used for each of the employed methods. Moreover, we have excluded $C_{gw}$ from our analysis assuming the hypothesis that the data consists of noise and a single CGW source.

The model parameters in equation (\ref{Eq:lik}) are split in two groups (i) parameters describing the CGW signal ($\vec{\lambda}: \vec{r} = \vec{r}(\vec{\lambda})$), and (ii) parameters describing the noise model $\vec{\theta}$. The waveform of a non-spinning circular binary given by equation (\ref{Eq:signal}) is generally described by $7+2N$ parameters. The Earth term (a single sinusoidal GW) is fully described by 7 parameters: ${(\mathcal{A},\theta_S,\phi_S,\Psi,\iota,\omega,\Phi_0)}$, and each pulsar term adds two additional parameters: frequency and phase ${(\omega_a,\Phi_a)}$. As mentioned above, the pulsar term might fall at the same frequency as the Earth term, in which case we have only one extra parameter per pulsar (since $\omega_a=\omega$). In principle, even for $\omega_a\neq\omega$, $\omega_a$ and $\Phi_a$ are uniquely connected through the pulsar distance $L_a$  as shown by equations (\ref{Eq:freqev}) and (\ref{Eq:phaseev}). However this implicitly assumes that we have an exact model for the evolution of the binary, which in this case is perfectly circular, non-spinning, and GW-driven. Even tiny deviations from these assumptions (i.e., small residual eccentricity, partial coupling with the environment, spins), very likely to occur in nature, will invalidate the $\omega_a-\Phi_a$ connection, and in the most general case, the two parameters must be considered separately. We will specify the details of the adopted waveform model for each individual method in the next section.  

Some of the noise parameters (like pulsar-intrinsic red noise or clock
and ephemeris errors) are correlated with the GW parameters and one
should in principle fit for noise and GW parameters
simultaneously. However, such a multidimensional search is
computationally very expensive, and in most of the searches detailed
below, we use noise characteristics derived from the SPA introduced in
Section~\ref{subsec:dataset} and extensively described in
\cite{EPTA2DM, EPTA3Noises}. We characterise the noise by considering
the data from each pulsar separately (as given by the SPA), and to
exclude any potential bias we also considered a model ``noise +
monochromatic signal''. The results of the latter are usually
consistent with the ``noise only'' model, except for one pulsar
, J1713+0747, where we have found some correlation between the
parameters describing the red noise and the extra monochromatic
signal. Each SPA returns posterior distributions for the array of
noise parameters $\vec{\theta}_a$: slope and amplitude of the red
noise, slope and amplitude of DM variations (both red noise and DM
variation were modeled as single power laws) and EFAC and EQUAD for
each backend. We have used this information in two ways: (i) use the
ML estimator for all the parameters ( $\vec{\theta}_{ML}=\cup
\vec{\theta}_{ML,a}$ ) and assume that the noise is represented by
that model, (ii) draw the parameters describing the noise from the
posterior distributions obtained in the SPA. The first choice (fixed
noise at $\vec{\theta}_{ML}$) is computationally cheaper as we need to
compute the noise variance matrix, $C(\vec{\theta})$, only once, while
in the second choice we need to recompute it for each draw of the
noise parameters. \cite{2014ApJ...794..141A} found that fixing the
noise to the ML values will bias the results of the search and will
yield a better upper limit compared to the full search including noise
parameters. The effect is, however, not dramatic, as they found their
upper limit is less than a factor $\approx1.5$ worse in the latter
case, over the full frequency range. To test the robustness of our
results we also ran a full analysis searching simultaneously on the GW
and noise parameters on a restricted dataset of 6 pulsars (see Section
\ref{subsec:RankPulsar} for the definition of this restricted
dataset).

\section{Methods}
\label{sec:methods}

As described in the previous section, the two parts of the GW signal
(Earth term and pulsar term) might or might not fall at the same
frequency, which has implications for the form of the likelihood
function given in equation (\ref{Eq:lik}). On top of that, both
frequentist and Bayesian methods can be used to analyze the data. We
therefore identify four separate classes of analysis: frequentist non-evolving, 
frequentist evolving, Bayesian non-evolving, and
Bayesian evolving. In each individual case, the signal (and sometimes
also the noise) can be treated differently, and the likelihood might
undergo peculiar manipulations (maximization, marginalization, etc.);
moreover, we sometimes explore two variations of the same method. The
result is a variety of complementary searches, which we now describe
in detail and which are comparatively summarized in Table \ref{tab2}.

\subsection{Frequentist methods}
\label{subsec:FreqAnalysis}

The basis for the frequentist analysis is to postulate that we have a
deterministic signal in the data which is either corrupted (in case of
detection) or dominated by noise (no detection). We then define
appropriate statistical distributions (or simply ``statistics'') based
on the likelihood function, those describe the data in absence and
presence of a signal. Those statistics must be chosen in such a way
that the detection rate is maximised for a fixed value of the false
alarm probability (FAP), which is also known as the Neyman-Pearson
criterion. The aim is to check the null hypothesis (i.e., whether the
data are described by noise only), and in case of ``no detection'' set
an upper limit on the GW amplitude. In building our statistics we
always assume that the noise is Gaussian and, unless otherwise stated,
characterised by ML parameters estimated during the SPA. We then
fix the FAP at 1\% to set the detection threshold. In case of no
detection, frequency dependent upper limits are obtained by splitting
the frequency range in small bins, 
\cR{performing a large number of signal injections}
with varying amplitude in each bin, and computing the
associated detection statistics. In the next two subsections we
describe two particular implementations of this procedure, known as
$\mF_p$ and $\mF_e$ statistics.

\subsubsection{$\mF_p$-statistic}
In the case of non-evolving sources (i.e. $\omega_a=\omega$), ~\cite{2012ApJ...753...96E} showed that, for each pulsar $a$, one can write equation (\ref{Eq:signal}) factorizing out the $\omega$ dependence
\begin{equation}
r_{a}(t) = \sum_{j=1}^2 b_{(j,a)}(\mathcal{A},\theta_S,\phi_S,\Psi,\iota,\Phi_0,\Phi_a) \kappa_{(j)}(\omega,t).
\end{equation}
Explicit expressions for $b_{(j,a)}$ and $\kappa_{(j)}$ can be found in ~\cite{2012ApJ...753...96E}. We merely stress here that $\kappa_{(j)}$ are independent on the considered pulsar $a$. The log of expression (\ref{Eq:lik}) can then be  maximized over the constant $2N$ coefficients  $b_{(1,a)},b_{(2,a)}$ assuming that they are independent, resulting in what is commonly known as the $\mF_p$-statistic. This latter assumption is not actually true if the number of pulsars is larger than 6. We make use of the full 41 pulsar dataset in the $\mF_p$-statistic evaluation. However, 6 pulsars dominate (i.e., give 90\% of) the S/N, as discussed in Section~\ref{subsec:RankPulsar}; moreover, we can simply postulate this form of statistic. Assuming independence in the maximisation process will somewhat degrade the detection power of the $\mF_p$-statistic; nonetheless, this is a very simple detection statistic which depends only on one parameter: the frequency of the GW. The $\mF_p$-statistic is in essence an ``excess-power'' statistic, in which we basically search for extra power --compared to the expectations of the statistic describing the null hypothesis-- at a given frequency in each pulsar's data (summing the weighted square of the Fourier amplitudes in each pulsar). By mixing the GW phases $\Phi_0$ and $\Phi_a$ at the Earth and at the pulsar in the coefficients $b_{(j,a)}$, by construction $\mF_p$ assumes that there is no coherence between GW signals across each individual pulsar's data. 
 
With the assumption of Gaussian noise, $2\mF_p$ follows a central $\chi^2$ distribution, $p_0(\mF_p)$ -- non-central, $p_1(\mF_p,\rho)$, if a signal with optimal S/N $=\rho$ is present -- with  $n=2N$ degrees of freedom (which follows from $2N$ maximisations of the log likelihood), given by:
\be
p_0(\mF_p) = \frac{\mF_p^{n/2-1}}{(n/2-1)!} {\rm exp}(-\mF_p),
\label{Eq:fp0}
\ee
\be
p_1(\mF_p,\rho) = \frac{(2\mF_p)^{(n/2-1)/2}}{\rho^{n/2-1}}I_{n/2-1}(\rho\sqrt{2\mF_p}) e^{-\mF_p-\frac{1}{2}\rho},
\label{Eq:fp1}
\ee
where $I_{n/2-1}(x)$ is the modified Bessel function of the first kind of order $n/2-1$. We divide our dataset in bins $\Delta f = 1/T$, where $T$ is the longest pulsar observation time, and we evaluate $\mF_p$ independently in each bin. At any given $f$, we consider only pulsars with observation time $T>1/f$ when evaluating $\mF_p$. Since the implementation of the $\mF_p$-statistic is computationally cheap, we run two flavors of it: (i) one in which we fix the noise to the ML values ($\mF_p${\it-ML} hereinafter), (ii) one in which we take into account the uncertainty in the noise parameters by sampling from their posterior distribution derived from the SPA (simply $\mF_p$ hereinafter).

\subsubsection{$\mF_e$-statistic}
\label{Method:Fe}
If the source is evolving, then $\omega_a\neq\omega$ and one can choose to consider as ``signal'' only the Earth portion of equation (\ref{Eq:signal}), and treat the pulsar term as a source of noise. In this case, we take only the Earth term of $\vec{r}$ in the likelihood expression  (\ref{Eq:lik}). \cite{2012PhRvD..85d4034B} showed that it  is convenient to rewrite the antenna pattern expressions (\ref{Eq:pattern}) separating the terms containing the polarization angle $\psi$:
\bea
F_{a}^{+} & =&  F^{a}_c \cos( 2\psi) + F^{a}_s \sin(2\psi) \\
F_{a}^{\times} & = & -F^{a}_s \cos(2\psi) + F^{a}_c \sin(2\psi),
\ena
thus rearranging equation (\ref{Eq:signal}) in the form 
\begin{equation}
r_{a}(t) = \sum_{j=1}^4 a_{(j)}({\mathcal{A},\iota,\psi,\Phi_0}) h_{(j)}^{a}(\omega,\theta_S,\phi_S,t).
\end{equation}
Explicit expressions for $a_{(j)},h_{(j)}^{a}, F^{a}_c, F^{a}_s$ can be found in \cite{2012PhRvD..85d4034B}, but note that, contrary to the $\mF_p$ case, the coefficients $a_{(j)}$ do not depend on the considered pulsar $a$. As done for the $\mF_p$ case, we can now maximize over the four constants $a_{(i)}$, constructing a statistic -- the $\mF_e$-statistic -- which is a function of three parameters; namely the GW frequency, $f$, and source sky location, $\theta_S, \phi_S$. If the noise is Gaussian, $\mF_e$ follows a $\chi^2$ distribution, $p_0(\mF_e)$ -- non-central, $p_1(\mF_e,\rho)$, if a signal with optimal S/N$=\rho$ is present -- with four degrees of freedom (which follows from the four maximisations of the log likelihood), given by:
\begin{equation}
p_0(\mathcal{F}_e)=\mathcal{F}_e e^{-\mathcal{F}_e},
\label{Eq:fe0}
\end{equation}
\begin{equation}
p_1(\mathcal{F}_e,\rho)=\frac{(2\mathcal{F}_e)^{1/2}}{\rho}I_1(\rho\sqrt{2\mathcal{F}_e})e^{-\mathcal{F}_e-\frac{1}{2}\rho^2},
\label{Eq:fe1}
\end{equation}
Where $I_1(x)$ is the modified Bessel function of the first kind of order 1.

Note that the $\mF_e$-statistic adds the data from the pulsars
coherently (taking the phase information into account), and being a
function of the source position, it allows direct sky
localization. However, if the signal is non-evolving, its efficiency
drops significantly and the estimation of the source sky location can
be severely biased. The evaluation of $\mF_e$ is significantly more
computationally expensive than $\mF_p$, as we need to take into
account cross-pulsar correlations. We therefore continue to use the
full 41-pulsar dataset, but we fix the noise parameters at the ML
values found in the SPA, without attempting any sampling (contrary to
the $\mF_p$ case).

For searching over the 3-D parameter space ($f$, $\theta$, $\phi$), we
use the multi-modal genetic algorithm described in
\cite{petiteau13}. A detailed description of the method can be found
there; here we give only a brief overview. In the first step we run a
genetic algorithm (with 64 organisms) over 1000 generations tuned for
an efficient exploration of the whole parameter space. Then we
identify the best spot in the 3-D space and seed there a variation of
the Markov chain Monte Carlo "MCMC Hammer"~\citep{mcmchammer}, which
serves as a sampler and returns the effective size of the ``mode'' and
the correlations among parameters. Here ``mode'' stands for a local
maximum of the likelihood. We then again run the genetic algorithm,
exploring the remaining parameter space (i.e. excluding the mode we
already found) and searching for other potential local maxima in the
likelihood. At each distinct maximum that is found, we run "MCMC
hammer", and we iterate this procedure 5 times. The end result is a
set of independent local maxima in likelihood, among which we choose
the largest (highest in value). We repeat the entire procedure three
times to verify convergence. This algorithm has proven to give very
robust results, even in the case of pathological likelihood surfaces
with multiple maxima.

\begin{table*}
\begin{center}
\begin{tabular}{c|ccccc}
\hline
Search ID & Noise treatment & pulsars ($N$) & parameters & Signal model & Likelihood\\
\hline
{\it Fp-ML}             & Fixed ML & 41 & 1 & E$+$P NoEv &Maximization over 2$N$ constant amplitudes\\ 
{\it Fp}             & Sampling posterior & 41 & 1 & E$+$P NoEv & Maximization over  2$N$ constant amplitudes\\ 
{\it Fe}             & Fixed ML & 41 & 3 & E & Maximization over 4 constant amplitudes \\ 
{\it Bayes\_E}        & Fixed ML & 41 & 7 & E & Full\\ 
{\it Bayes\_EP}       & Fixed ML & 6 & $7+2\times6$ & E$+$P Ev & Full\\ 
{\it Bayes\_EP\_NoEv}  & Fixed ML & 41 & 7 & E$+$P NoEv & Marginalization over pulsar phases\\ 
{\it Bayes\_EP\_NoEv\_noise}  & Searched over & 6 & $7+5\times6$ & E$+$P NoEv &  Marginalization over pulsar phases\\ 
\hline
\end{tabular}
\end{center}
\caption{Summary of the searches performed in this study. Column 1: name of the search; column 2: treatment of the noise in the search; column 3: number of pulsars considered in the dataset ($N$); column 4: dimensionality of the parameter space to search over; column 5: adopted signal model; column 6: notes about the treatment of the likelihood function. The different signal models are: Earth term only (E), Earth plus pulsar term at the the same frequency (E$+$P NoEv), Earth plus pulsar term at different frequencies (E$+$P Ev).}
\label{tab2}
\end{table*}

\subsection{Bayesian analysis}
\label{subsec:BayesAnalysis}
In the Bayesian approach, the parameters describing the model are treated as random variables. The principles of Bayesian statistics provide a robust framework to obtain probability distributions of those parameters for a given set of observations, while also folding in our prior knowledge of them.

Bayes' theorem states that the \textit{posterior} probability density function (PDF), $p(\vec{\mu}|\vec{d},\mathcal{H})$, of the parameters $\vec{\mu}$ within a hypothesised model $\mathcal{H}$ and given data $\vec{d}$ is
\begin{equation} \label{eq:bayes-theorem}
p(\vec{\mu}|\vec{d},\mathcal{H}) = \frac{\mathcal{L}(\vec{\mu},\mathcal{H}|d)\pi(\vec{\mu}|\mathcal{H})}{p(\vec{d}|\mathcal{H})},
\end{equation}
where $\mathcal{L}(\vec{\mu},\mathcal{H}| \vec{d})$ is the likelihood of the parameters given the data assuming the model $\mathcal{H}$ with parameters $\vec\mu$. The prior PDF of the model parameters, $\pi(\vec{\mu}|\mathcal{H})$, incorporates our preconceptions and prior experience of the parameter space. The Bayesian evidence, $p(\vec{d}|\mathcal{H})\equiv\mathcal{Z}$, is the probability of the observed data given the model $\mathcal{H}$,
\begin{equation}\label{eq:evidence}
\mathcal{Z} = \int \mathcal{L}(\vec\mu)\pi(\vec\mu)d^N\hspace{-2 pt}\mu.
\end{equation}
For posterior inference within a model, ${\cal Z}$ plays the role of a normalisation constant and can be ignored. However, if we want to perform model selection then the evidence value becomes a key component in the computation of the posterior odds ratio:
\begin{equation}
\frac{p(\mathcal{H}_2|\vec d)}{p(\mathcal{H}_1|\vec d)} = \frac{p(\vec d|\mathcal{H}_2)p(\mathcal{H}_2)}{p(\vec d|\mathcal{H}_1)p(\mathcal{H}_1)}=\frac{\mathcal{Z}_2\times p(\mathcal{H}_2)}{\mathcal{Z}_1\times p(\mathcal{H}_1)}.
\label{Eq:odds}
\end{equation}
Here $\mathcal{H}_1,\mathcal{H}_2$ are the two models under comparison, $\mathcal{Z}_2/\mathcal{Z}_1$ is the Bayes factor, and $p(\mathcal{H}_2)/p(\mathcal{H}_1)$ is the prior probability ratio for the two competing models, which is often set to be one; the posterior odds ratio is then just the Bayes factor.

When we specialize the formalism above to PTA data, the likelihood
$\mathcal{L}$ is given by equation (\ref{Eq:lik}), the data $\vec d$
is the concatenation of the TOA series $\vec{\delta t}$ and the
model's parameters $\vec{\mu}$ are identified with $\vec{\theta},
\vec{\lambda}$. The presence of a signal in the data is assessed
through the odds ratio, equation (\ref{Eq:odds}), where model
$\mathcal{H}_1$ corresponds to data described by noise only and model
$\mathcal{H}_2$ corresponds to data described by noise plus a CGW.  We
use non-informative and conservative priors in our analysis: they are
always uniform in all parameters. It is especially important to
emphasise that we have used a broad uniform prior in the signal's
amplitude $\mathcal{A}$, which will provide robust and conservative
upper limits on the strain. Unless otherwise stated, we fix the
stochastic noise parameters to the ML values found in the SPA, and we
employ either MultiNest \citep{MULTINEST} or a parallel tempering
adaptive MCMC \citep{Lassus15} to return samples from the posterior
distributions, and thus reconstruct the posterior PDFs. Both
techniques also permit a recovery of the aforementioned
model-comparison statistic known as the \textit{Bayesian evidence}. We
describe the different searches in detail below.

\subsubsection{Phase-marginalised Bayesian analysis} 

For the non-evolving CGW signal searches, one should sample a $7+N$
multidimensional posterior, corresponding to the parameters
${(\mathcal{A},\theta_S,\phi_S,\Psi,\iota,\omega,\Phi_0,\Phi_a)}$. For
a 41-pulsar dataset, that amounts to sampling a 48-dimensional
parameter space. We mitigate the computational cost implied by such
high-dimensionality by numerically marginalising (integrating) our
posterior distribution over all of these nuisance parameters $\Phi_a$,
thereby collapsing the search to more manageable dimensions
\citep[i.e. to seven parameters only,][]{taylorellisgair14}. By doing
so, we not only rapidly recover the posterior PDFs, but also
achieve quick and accurate Bayesian evidence values. This method is
close in spirit to the $\mF_p$-statistic, we call it
\textit{phase-marginalised Bayesian analysis} (labeled as {\it
  Bayes-EP-NoEv}, for Bayes, Earth+pulsar, non-evolving source). The
sampling of the posterior is performed by MultiNest.

\begin{figure*}
\includegraphics[width=\textwidth]{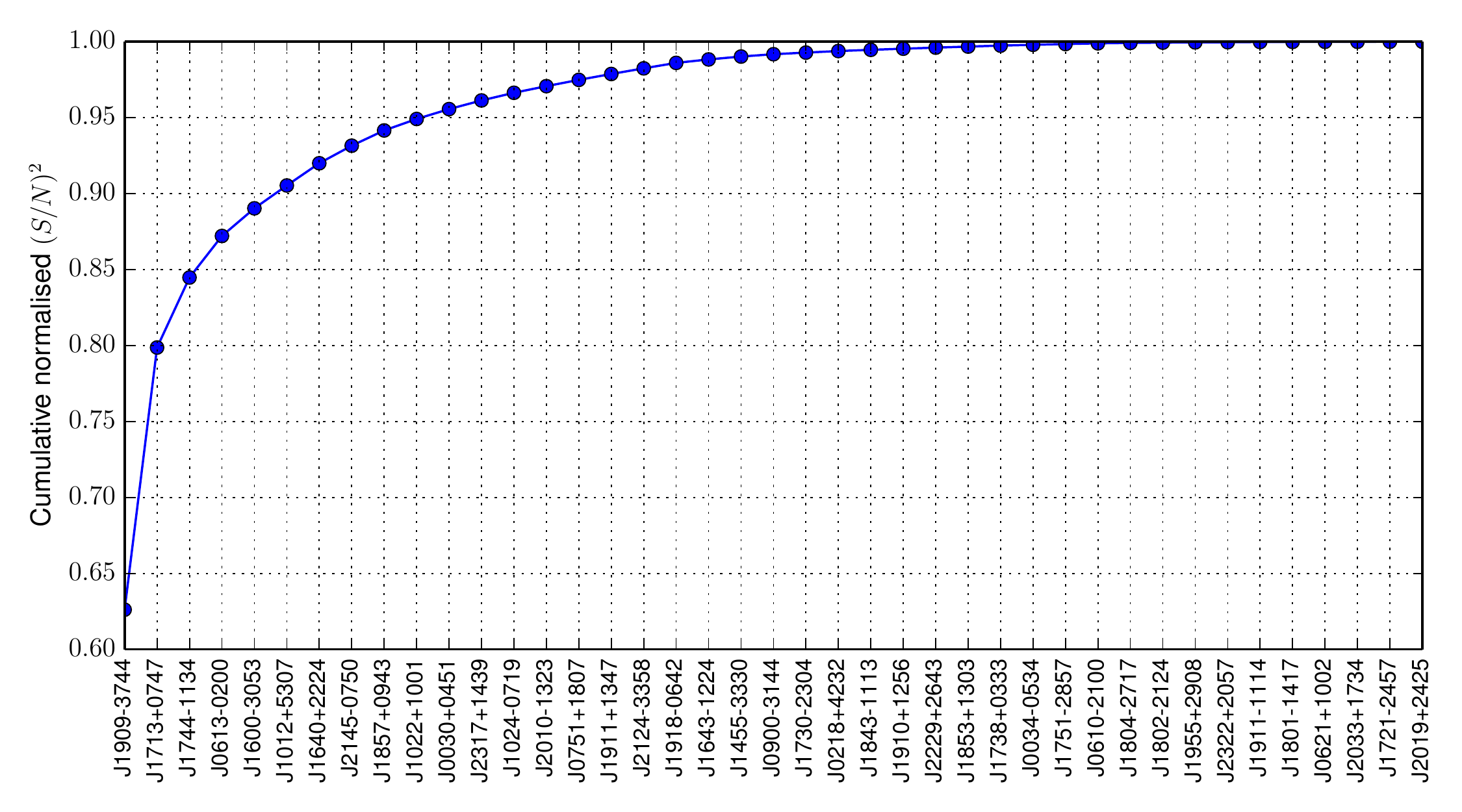}
\caption{Cumulative normalised $(\text{S/N})^2$. We rank pulsars according to their relative contribution to the total S/N, and we plot the quantity $\sum_{i<a}\rho_a/\rho$.}
\label{fig:snr_contr}
\end{figure*}

The high performance of the MultiNest sampler allowed us to also run a
full search, including noise parameters $\vec{\theta}$, on a restricted
dataset composed of the 6 best pulsars --contributing to 90\% of the
total S/N, see Section \ref{subsec:RankPulsar} and figure
\ref{fig:snr_contr}-- in our array (labeled {\it
  Bayes-EP-NoEv-noise}). We use non-informative priors also for the
pulsar noise parameters: uniform in the range [1,7] for the slopes of
both red noise and DM; uniform in the range [$-20, -10$] for the
logarithm (base 10) of their amplitudes; uniform in the range [0,10]
for the global EFAC \citep[see][for a definition of global EFAC]{EPTA4isotropic}. The posterior spans now a $7+5N=37$ dimension parameter
space. We designed this search as a benchmark for our different
fixed-noise analysis; to speed up the sampling, we restricted the
frequency prior range to [5,15]nHz, which turns out to be the
sweet-spot of our array sensitivity.

\subsubsection{Full Bayesian analysis}

We also employed the Bayesian formalism to construct a search for
evolving signals. In this case we can either (i) use the full signal
of equation (\ref{Eq:signal}), or (ii) restrict ourselves to the Earth
term only. This latter case (ii) is similar in spirit to the
frequentist $\mF_e$-statistic, with the difference that we now search
over the full 7-dimensional source parameter space with the whole
41-pulsar array; we label this analysis {\it Bayes-E}. The full-signal
analysis (i) is very computationally expensive because we do not
assume any model for the source evolution, therefore adding two extra
parameters ${(\omega_a,\Phi_a)}$ for each pulsar. \cR{Note that our
parametrisation of the full signal does not rely on any knowledge
about the distance to the pulsar.} We name this search
{\it Bayes-EP-Ev}. We use non-informative priors also on those
parameters, imposing the only constraint that the pulsar-term
frequencies cannot be higher than the Earth-term one ($\omega_a\le
\omega$). Since we need to cover a $7+2N$ parameter space, we limit
this search to the best 6 pulsars, for a total of 19 parameters. In
both searches, the multidimensional posteriors are stochastically
sampled with a parallel tempering adaptive MCMC \citep{Lassus15}.

\subsection{Restricted dataset: ranking pulsar residuals}
\label{subsec:RankPulsar}
As mentioned above, some of the searches are extremely computationally expensive, involving sampling of highly dimensional parameter spaces. A way to mitigate the computational cost is to run the algorithms on a ``restricted dataset'', which includes only the best pulsars for these purposes. We therefore need a metric to rank the quality of each pulsar. We choose our metric to be the relative S/N contribution of each pulsar to a putative detectable source.  We conduct extensive Monte Carlo simulations, in which we inject CGW signals with an astrophysically motivated distribution of parameters $\vec{\lambda}$ in the EPTA dataset. For each signal we compute the total S/N and the relative contribution of each individual pulsar according to the standard inner product definition:  
\begin{equation}
(\text{S/N})^2  = (h(\vec{\lambda)} | h(\vec{\lambda})) \equiv (h(\vec{\lambda})^T G) (G^T C(\vec{\theta}) G)^{-1} (G^Th(\vec{\lambda})).
\end{equation}
For each injected CGW, we draw the noise parameters of each pulsar
($\vec{\theta}_a$) from the corresponding posterior distribution of
the SPA. We injected 1000 detectable CGWs, and each signal contained
both Earth and pulsar terms. In figure~\ref{fig:snr_contr} we plot the
average (over the 1000 simulations performed) build-up of $\rho =
(\text{S/N})^2$ as we add pulsars to the array, from the best to the
worst. The plot shows the cumulative S/N square over the total S/N
square, i.e., $\sum_{i<a}\rho_a/\rho$. The ranking reveals that the
array is heavily dominated by PSR J1909$-$3744, contributing more than
60\% of $\rho$, followed by PSR J1713+0747 at about 20\%. We decided
to form a dataset considering only the best pulsars building up 90\%
of $\rho$. We ended-up with 6 pulsars (PSRs J1909$-$3744, J1713+0747,
J1744$-$1134, J0613$-$0200, J1600$-$3053, and J1012+5307) which
constitute the restricted dataset mentioned above, { and is the same used in our companion isotropic and anisotropic GW background searches \citep{EPTA4isotropic, thgm15}}.

\section{Results}
\label{sec:results}

We now turn to the description of the main results of our analysis. As
in the previous section, we present the outcome for the frequentist
and Bayesian analyses separately, discussing first the detection
results and then the upper limit computation for the various
techniques. No evidence for CGW signals was found
in the data; a summary of all the 95\% upper limits on the GW strain
amplitude $\mathcal{A}$ as a function of frequency is collected in
figure \ref{fig:upperlimit_all}.

\subsection{Frequentist analysis}
\subsubsection{$\mF_p$-statistic}
\label{subsec:resultsFp}

\begin{figure}
\includegraphics[width=\columnwidth]{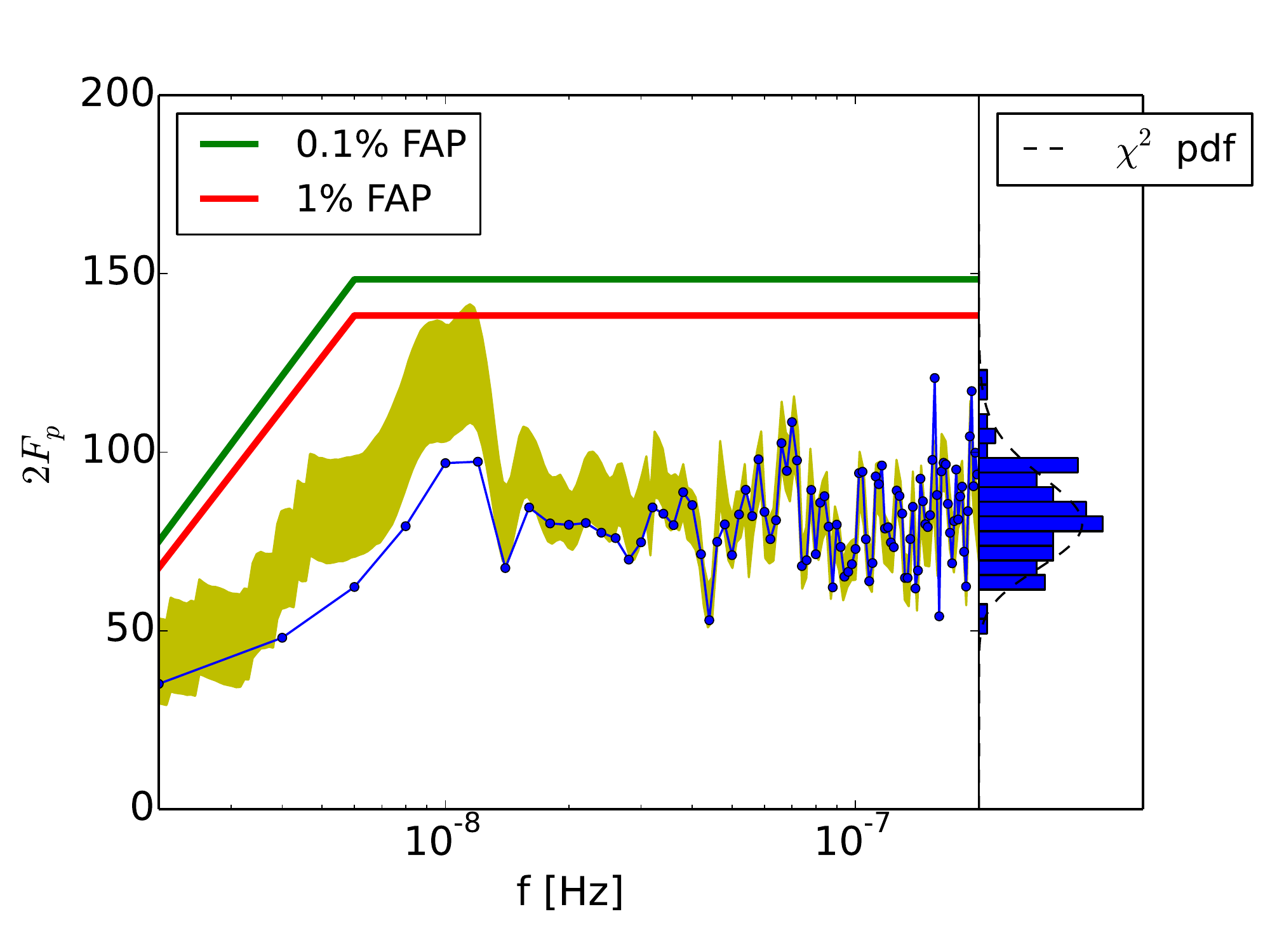}
\caption{$\mF_p$-statistics. The blue line represents $\mF_p$ evaluated at 99 independent frequencies, for 41 pulsars, using the ML noise parameters. \cR{The right panel shows the histogram of the $\mF_p$ values at all frequencies, and the dashed line is a central $\chi^2$-distribution, which is the expected distribution of the $\mF_p$ statistic in absence of a GW signal. The two are consistent with a p-value of $0.93$, which is indicative of excellent agreement.} The yellow area denotes the central $90\%$ of $\mF_p$ evaluated across the whole sample of noise parameters. The thresholds turn over below 6 nHz because of the reduced number of pulsars that have enough observing
  time span.}
\label{fig:Fpnondetection}
\end{figure}

To tackle the issue of detection, we have evaluated $\mF_p$ at $N_f=99$ independent frequencies in the range $\left[ 1/T, 2\times10^{-7} {\rm Hz} \right]$ in steps of $\Delta f = 1/T$, where $T$ is the maximum observation time. In this dataset $1/T = 2.0 \times 10^{-9} {\rm Hz}$. In the absence of a signal,  $2\mF_p$ follows the $\chi^2$ PDF with $n=2N$ degrees of freedom given by equation (\ref{Eq:fp0}). The false alarm probability (FAP) associated to a given threshold $F_0$ is simply given by the integral of the PDF and takes the form: 
\be
P(\mF_p > F_0) = \int_{F_0}^{\infty} p_0(\mF_p) d\mF_p = {\rm exp}(-\mF_0) \sum_{k = 0}^{n/2-1} \frac{F_0^k}{k!}.
\ee
To assess the global probability of finding a given value of $\mF_p$ in our dataset, we need to take into account that we are evaluating it at 99 different frequency bins, i.e., we are performing 99 independent trials. The global FAP is therefore given by
\be
{\rm FAP} = 1 - \left[ 1 - P(\mF_p > F_0) \right]^{N_f}.
\label{Eq:fap}   
\ee
We choose thresholds in $F_0$ that correspond to FAPs of $0.01$ and $0.001$. The results are presented in figure ~\ref{fig:Fpnondetection}. For the ML noise parameters (solid line with circles and corresponding histogram on the right),  the data is consistent with the noise description with a p-value of $0.93$ and that there is no excess at any frequency above the $0.01$ FAP threshold. However, the choice of the noise parameters, and hence the covariance matrix $C$, is crucial in evaluating $\mF_p$, and the SPA reveals that many parameters are poorly constrained. We therefore chose to create a whole distribution of $\mF_p$ at each frequency, sampling over noise parameters from the posterior distributions produced by the SPA. This is overplotted as the yellow area in figure~\ref{fig:Fpnondetection}. The results obtained in this way are independent of the particular ML value and give a better view of the uncertainties involved. At each frequency, $f$, only pulsars with baselines $T > 1/f$ have been taken into account. This explains the rising FAP thresholds. The uncertainty in $\mF_p$ induced by the uncertainty in the noise parameters is much larger at lower frequencies, where red noise and DM dominate the noise model. For the lowest frequencies, the ML evaluation of $\mF_p$ does not even lie within the $90\%$ region shown in the plot, which is a consequence of the fact that we sum the contributions of 41 broad, and often skewed, distributions. In several pulsars the ML red noise or DM amplitudes tend to lie at the upper end of the distribution, which leads to small values of $\mF_p$ and results in this offset. 

After confirming that the SPA describes the data appropriately as only
noise, we want to know how large a CGW contribution must be in order
to make the $\mF_p$-distribution non-central and clearly
distinguishable from the noise. This yields an upper limit on the GW
strain $\mathcal{A}$ that might be present in our data and still
consistent with the observed $\mF_p$ value. A standard way to do this
in frequentist analysis is through signal injections. We first fix the
noise to the SPA-ML value; in this case the procedure to obtain the
upper limit on $\mathcal{A}$ at each frequency $f$ is as follows
\citep[see, e.g., ][]{2012ApJ...753...96E}:

\begin{enumerate}
\item compute $\mF_{p,0}$ using the dataset;
\item create 1000 different mock datasets $i$, by injecting in each of them one source with fixed strain $\mathcal{A}$ but otherwise random parameters, and compute $\mF_{p,i}$;
\item compute the fraction $y$ of mock datasets where $\mF_{p,i} > \mF_{p,0}$;
\item repeat steps (ii) and (iii) with different $\mathcal{A}$ until $y = 0.95$.
\end{enumerate}

In practice, we iterate over a grid in ${\rm log_{10}}\mathcal{A}$ with
step size $0.1$ and interpolate to find the point for which $y =
0.95$. We also want to obtain an upper limit that takes into account
the uncertainty of the single-pulsar noise parameters. In doing this,
the procedure outlined above is modified as follows:

\begin{enumerate}
\item compute a distribution of 1000 $\mF_{p,0}$ at each frequency $f$ using the dataset and 1000 noise parameters drawn from the posterior noise PDF obtained from the SPA;
\item create 1000 different mock datasets $i$, by injecting in each of them one source with fixed strain $\mathcal{A}$ but otherwise random parameters in each, and compute $\mF_{p,i}$, each time drawing different noise parameters from the posterior noise PDF;
\item compute the fraction $y$ of mock datasets where $\mF_{p,i} > \bar{\mF}_{p,0}$, where the single reference $\bar{\mF}_{p,0}$ is chosen to be the mean of the distribution obtained in step (i);
\item repeat steps (ii) and (iii) with different $\mathcal{A}$ until $y = 0.95$.
\end{enumerate}
In other words, we want 95\% of the $\mF_{p,i}$ distribution to lie
above the mean value $\bar{\mF}_{p,0}$ of the observed
distribution. The motivation behind criterion (iii) is that it
resembles the criterion for the ML upper limit and it is much more
conservative than a Kolmogorov-Smirnov test. It is also possible to
choose even more conservative criteria, susceptible to possible non-Gaussian tails in either distribution, but we do not explore this
possibility here.

\begin{figure}
\includegraphics[width=\columnwidth]{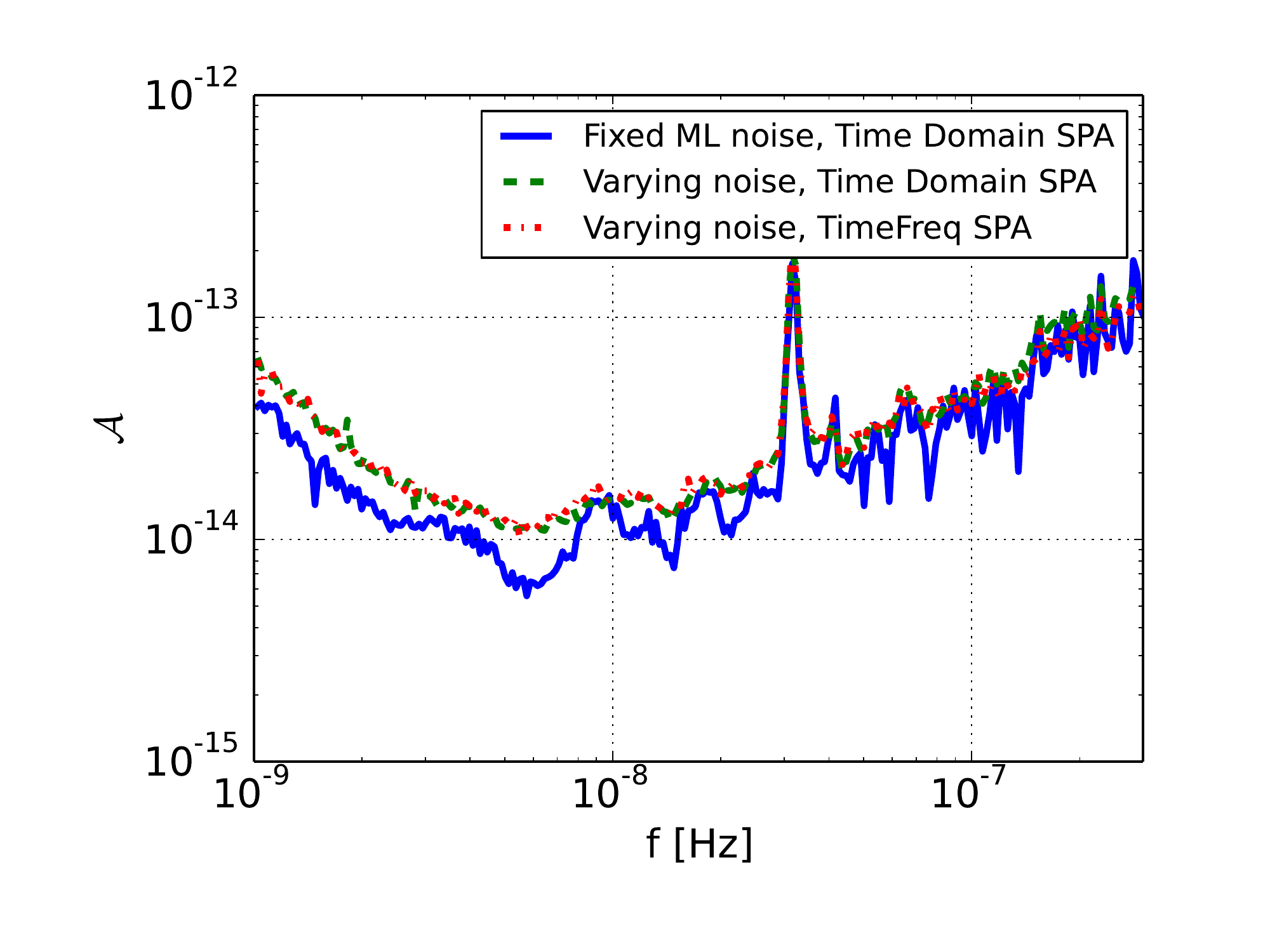}
\caption{$95\%$ upper limit on the GW strain $\mathcal{A}$ obtained with the $\mF_p$-statistic. The blue line corresponds to noise parameters fixed to the ML values obtained in the time-domain SPA, while the dashed lines take into account the full uncertainty in the noise estimation by sampling from the PDF distributions of either the pure time domain  (green) or the time-frequency (red) SPA. At 1/year (the peak), the limit is poor because the GW signal at this frequency is absorbed in the fitting of the pulsar positions.}
\label{fig:upperlimit}
\end{figure}

Upper limits as a function of $f$ are shown in figure
\ref{fig:upperlimit}. In all cases, the most stringent limit is around
$6-7$nHz, and reaches down to $\mathcal{A}=6\times10^{-15}$ for the
$\mF_p$-ML case. We note, however, that when we allow the noise to
vary, we get a limit which is a factor $\approx2$ worse at low
frequency, yielding a value of $\mathcal{A}=1.1\times10^{-14}$. This
is consistent with figure ~\ref{fig:Fpnondetection}; at low frequency
the ML estimator of the noise parameters is not representative of the
noise posterior distribution, resulting in $\mF_p$ values which are
biased  toward low values. Injections with lower $\mathcal{A}$ will therefore result
in a detectable excess $\mF_p$, pushing down the upper limit. Note
that our upper limit gets significantly worse at $f<3\times10^{-9}$Hz,
because red noise becomes significant for some pulsars, and not all 41 pulsars in
our array contribute down to those frequencies, having an observation
time $T<10$ years. At high frequency the 95\% upper limit degrades
approximately linearly with $f$, consistent with a white-noise-dominated dataset.

\subsubsection{$\mF_e$-statistic}
\label{sec:subsubFe}
\begin{figure}
\includegraphics[width=\columnwidth]{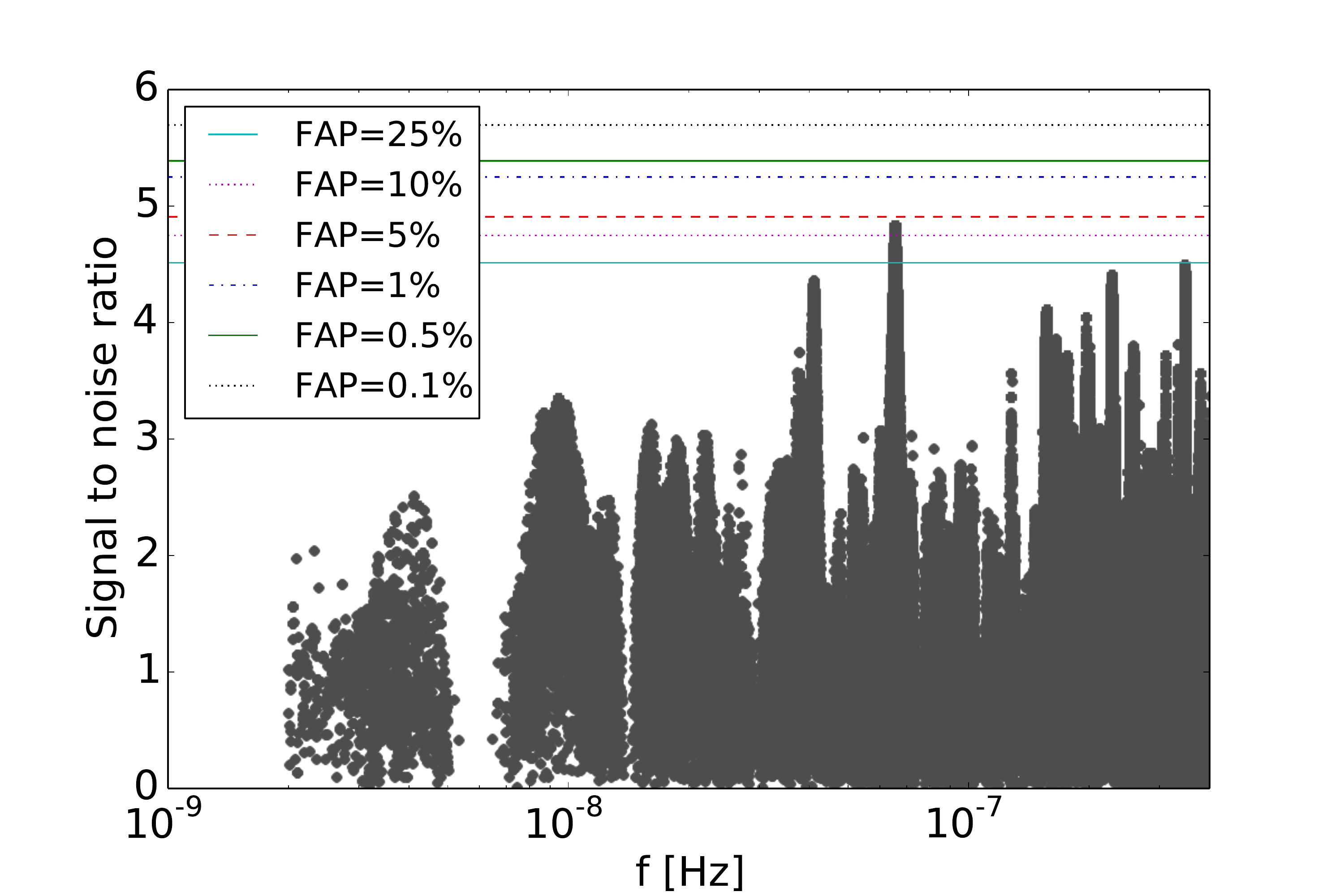}
\caption{ Result of the $\mF_e$-statistic detection analysis. The points are the trials of the search.  The horizontal lines are the detection thresholds for different FAPs.}
\label{fig:Fenondetection}
\end{figure}

Since $\mF_e$ is also a frequentist technique, the procedures to assess detection and to place upper limits are analogous to the $\mF_p$ case. Here, in absence of signal,  $2\mF_e$ follows the $\chi^2$ PDF with $n=4$ degrees of freedom given by equation (\ref{Eq:fe0}). The FAP associated with a given threshold $F_0$ is simply given by the integral of the PDF and takes the form: 
\be
P(\mF_e > F_0) = \int_{F_0}^{\infty} p_0(\mF_e) d\mF_e = (1+F_0)e^{-F_0}.
\ee
Again, the global false alarm rate depends on the number of trials, according to equation (\ref{Eq:fap}), which is now given by the number of independent templates in the sky location-frequency 3-D parameter space. The vast majority of templates we have used in the search are strongly correlated. We estimate the number of independent trials by constructing a stochastic 3-D template bank \citep[see][]{Babak:2008rb, Harry:2009ea}. We use a minimal match equal to 0.5 as the criterion of independence among different templates, and obtain 4276 independent points in the searched parameter space (full sky and frequency band restricted to $2-400$nHz). Figure~\ref{fig:Fenondetection} presents the result of the detection analysis. The maximum of $\mF_e$, $\mF_{e,{\rm max}}$, is found at $f$ = 66 nHz and $\{\theta_S,\phi_S\} = \{ 51.9^{\rm o} , 136.4^{\rm o}\} $. It corresponds to a FAP of 7\%, which is compatible with a non-detection. \cR{Note that a GW signal with S/N$=5$, if present in the data, would correspond to a FAP of $\approx 5$\%, which is clearly too high to make any confident claim of detection, as shown in figure~\ref{fig:Fenondetection}.}

Since the data are compatible with describing only noise, we can again compute the 95\% upper limit on the strain amplitude ${\mathcal A}$ of a putative CGW as a function of $f$ by means of signal injections. The procedure is similar to the one employed in the $\mF_p$-{\it ML} analysis. Here, we construct a grid of frequencies, and at each grid point we consider a small frequency interval $\Delta{f}$~=~1~nHz which is sufficient to capture the injected Earth term:
\begin{enumerate}
\item compute $\mF_{e,max,0}$, i.e. the maximum of $\mF_{e}$ on the whole sky and in the narrow frequency band $\Delta{f}$ in the raw dataset; 
\item create 1000
different mock datasets $i$, by injecting in each of them one source with fixed strain $\mathcal{A}$ but otherwise random parameters, including both Earth and pulsar terms; 
\item for each dataset $i$, run a search (stochastic bank + multi-search genetic algorithm{\footnote{A full use of the multi-modal genetic algorithm for each injection is computationally expensive and not needed, in practice. We therefore use a lighter and faster search for the injected signals. We construct a stochastic bank with minimal match 0.95 and we filter the data through this bank. We then identify the maximum of $\mF_e$ across the bank and refine our search running the genetic algorithm with 64 organisms evolved over 1000 generations. The stochastic bank is generated only once for the full parameter space and contains 532~488 templates. In each search we use only the portion of template bank covering the parameter space region around the injected signal.}}) to find $\mF_{e,max,i}$, i.e., the maximum of $\mF_{e}$ on the whole sky and in the narrow frequency band $\Delta{f}$; 
\item compute the fraction $y$ of the mock datasets in which $\mF_{e,max,i} > \mF_{e,max,0}$;
\item repeat steps (ii) and (iii) increasing  $\mathcal{A}$ until $y = 0.95$.
\end{enumerate}
The 95\% sky-averaged upper limit obtained in this way is shown by the red curve in figure~\ref{fig:upperlimit_all} and is in agreement with the results obtained with other methods.

\subsection{Bayesian analysis}

\begin{figure}
\includegraphics[width=\columnwidth]{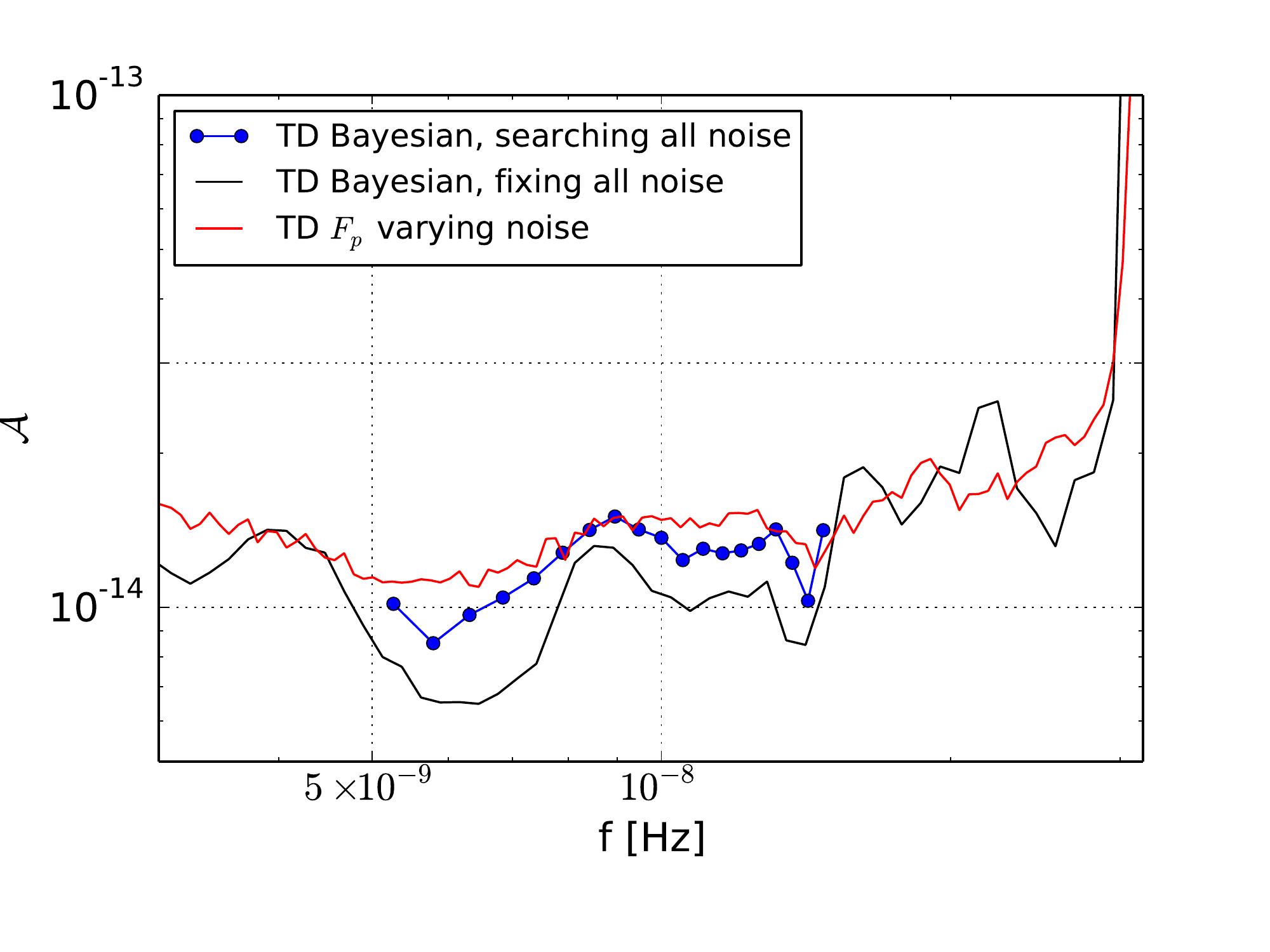}
\caption{The 95\% upper limit on the GW strain obtained with the phase-marginalised Bayesian analysis, searching on noise and signal simultaneously ({\it Bayes\_EP\_NoEv\_noise}, blue-circled curve), is compared to the same analysis with ML-fixed noise ({\it Bayes\_EP\_NoEv}, black curve) and to the noise-sampling $\mF_p$ analysis (red curve). \cR {Details of each specific analysis are given in Table~\ref{tab2}. "TD" stands for time-domain analysis as opposed to the time-frequency approach developed in \protect\cite{2013PhRvD..87j4021L}.}}
\label{fig:bayesnoise}
\end{figure}

\begin{figure*}
\includegraphics[width=\textwidth]{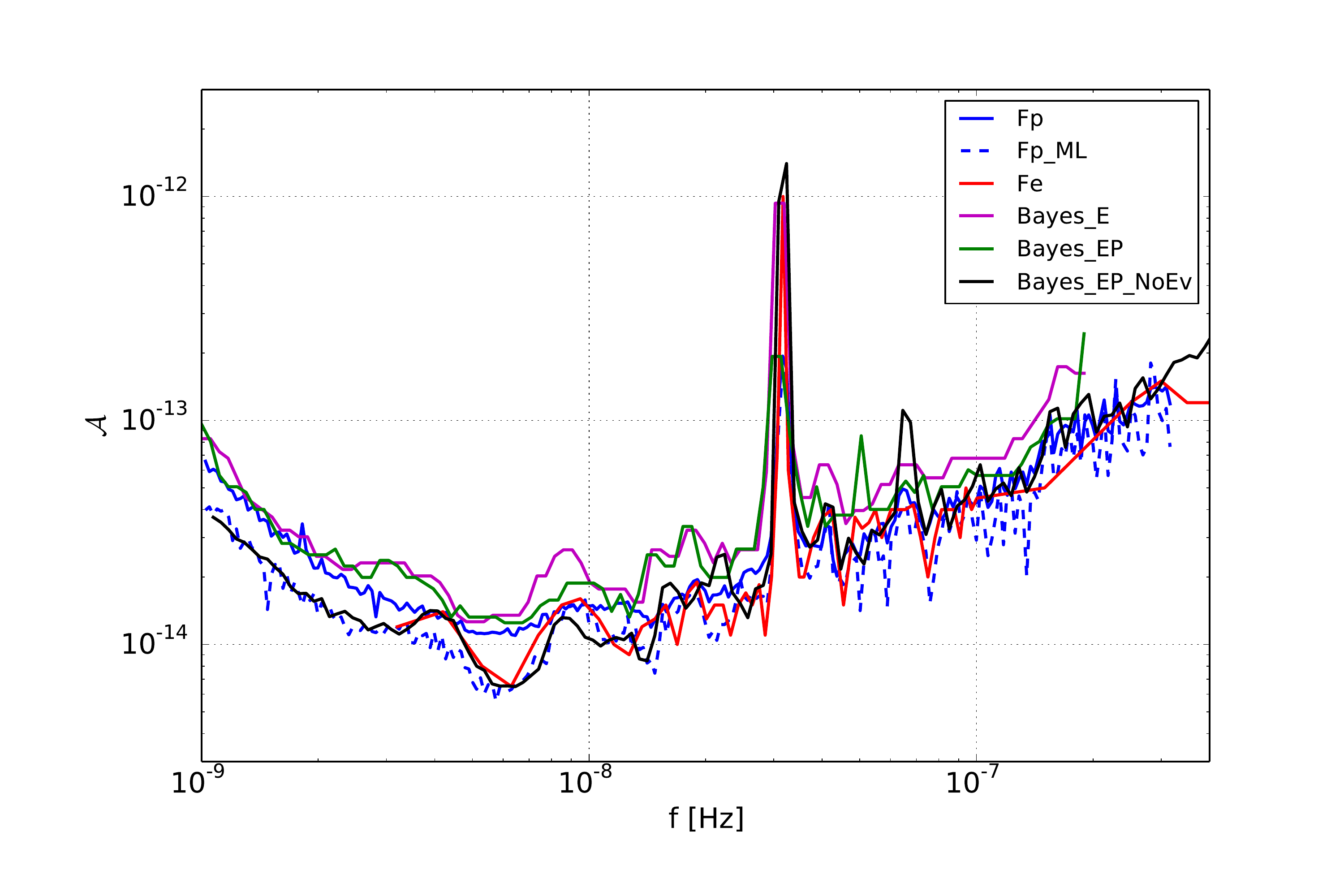}
\caption{The 95\% upper limit on the gravitational wave strain for the 3 frequentist methods, i.e. $\mF_p$ varying noise ({\it Fp}), $\mF_p$ fixed noise ({\it Fp\_ML}) and $\mF_e$, and the 3 bayesian methods, i.e. ``evolving source'' with Earth term only ({\it Bayes\_E}) and with Earth and Pulsar terms ({\it Bayes\_EP}) and ``non-evolving source'' with Earth and Pulsar terms ({\it Bayes\_EP\_NoEv}), see Table~\ref{tab2} for details.}
\label{fig:upperlimit_all}
\end{figure*}

The frequentist analysis presented above already provided strong evidence against the presence of a signal in the data. Nonetheless, this can also be addressed in the Bayesian framework through the computation of the odds ratio defined by equation (\ref{Eq:odds}). Since we give {\it a priori} no preference to the presence or absence of a signal in the data, we set the prior probability ratio to unity, and the odds ratio coincides with the Bayes factor. The Bayes factor is then simply the ratio of the Bayesian evidence computed for the hypothesis $\mathcal{H}_1$ and  hypothesis $\mathcal{H}_0$ as given by equation (\ref{Eq:odds}), which in our case reduces to:  
\begin{equation}
\mathcal{B}= \frac{\int \mathcal{L}( \vec{\theta}, \vec{\lambda}| \vec{\delta t} )\pi(\vec{\theta}, \vec{\lambda})d\vec{\theta}\,d\vec{\lambda}}{\int \mathcal{L}( \vec{\theta}| \vec{\delta t})\pi(\vec{\theta})d\vec{\theta}}.
\end{equation}
In the case of fixed noise, we assume that the noise parameters
$\vec{\theta}$ are known exactly (fixed at their ML value), and the
Bayes factor is directly computed from the likelihood ratio multiplied
by the priors, integrated over the source parameters
$\vec{\lambda}$. We compute the evidence using both MultiNest and
parallel tempering MCMC searches. In all the Bayesian searches with
fixed noise we obtain Bayes factors close to zero, consistent with a
non-detection and with the outcome of the frequentist analysis. In
particular, we get $\log(\mathcal{B})=-0.27$ for the search {\it Bayes\_E},
and $\log(\mathcal{B})=-0.31$ for the search {\it Bayes\_EP\_NoEv}.

The Bayesian analysis also returns samples from the joint posterior
probability distribution of all model parameters. The marginalized
distribution of any parameter of interest can then be evaluated by
integrating (i.e., marginalizing) the joint posterior distribution
over all other parameters. We are particularly interested in the
strain amplitude $\mathcal{A}$. We can then split the vector parameter
$\vec {\lambda}=(\mathcal{A},\vec{\lambda'})$ and integrate over
$\vec{\lambda'}$ to obtain the marginalized posterior for the
parameter $\mathcal{A}$. In practice, we divide the frequency range in
small bins in which we carry out this marginalization procedure
separately. The 95\% upper limit at each frequency corresponds to the
value $\mathcal{\tilde{A}}$ for which 95\% of the posterior
distribution lies at $\mathcal{A}<\mathcal{\tilde{A}}$; namely
\begin{equation}
0.95=\int_0^{\mathcal{\tilde{A}}}d\mathcal{A}\int d\vec{\lambda'} \mathcal{L}( \mathcal{A}, \vec{\lambda'} | \vec{\delta t}) \pi(\mathcal{A})\pi(\vec{\lambda'}).
\end{equation}

Results are shown in figure~\ref{fig:upperlimit_all}, which compares all the upper limits on the GW strain achieved by all methods presented in this paper. For the non-evolving source case, the {\it Bayes\_EP\_NoEv} upper limit agrees particularly well with the fixed-noise $\mF_p$-{\it ML} statistic. This is encouraging, since the two methods are similar in spirit as they adopt the same signal model and assume fixed/known noise parameters. For the evolving source case, the $\mF_e$ upper limit is very similar to both $\mF_p$-{\it ML} and {\it Bayes\_EP\_NoEv}, mimicking almost perfectly their behavior at low frequency.
The upper limits obtained by both the {\it Bayes\_E} and the {\it Bayes\_EP} searches are noisier and slightly higher, but overall consistent with the others within a factor of two.

As mentioned in section \ref{sec:methods}, we also ran a full
37-dimensional search over noise and signal parameters on the
restricted set of the 6 best pulsars in our PTA (c.f. figure
\ref{fig:snr_contr}), and in a restricted frequency range of
$5-15$nHz { where we have the best sensitivity}. We used the phase-marginalised Bayesian analysis for
non-evolving sources, and labeled the run {\it
  Bayes\_EP\_NoEv\_noise}. The 95\% upper limit obtained in this case
is shown in figure \ref{fig:bayesnoise}, together with the fixed noise
{\it Bayes\_EP\_NoEv} and the noise-sampling $\mF_p$ results. The {\it
  Bayes\_EP\_NoEv\_noise} limit lies a factor $1.1-1.5$ above the {\it
  Bayes\_EP\_NoEv} one. This is in line with the findings of
\cite{2014ApJ...794..141A}, and confirms that our ML fixed noise upper
limits are reliable within a factor $\lesssim1.5$. It is also
interesting to see that the {\it Bayes\_EP\_NoEv\_noise} limit agrees
fairly well with the noise-sampling $\mF_p$ one.  By analysing
figure~\ref{fig:upperlimit_all} we can conclude that all the upper
limits yielded by the different techniques agree within a factor of
two.  We also observe that methods based on fixed noise (ML)
parameters slightly underestimate the upper limit, which could be
because the ML values are not always representative of the posterior
distribution of the noise parameters.

\subsection{Sky maps}

Most of the searches outlined above are also sensitive to the source
location on the sky. We can therefore extend our study and produce sky
maps of the 95\% upper limits provided by our analysis as a function
of frequency.

\begin{figure}
\includegraphics[width=\columnwidth]{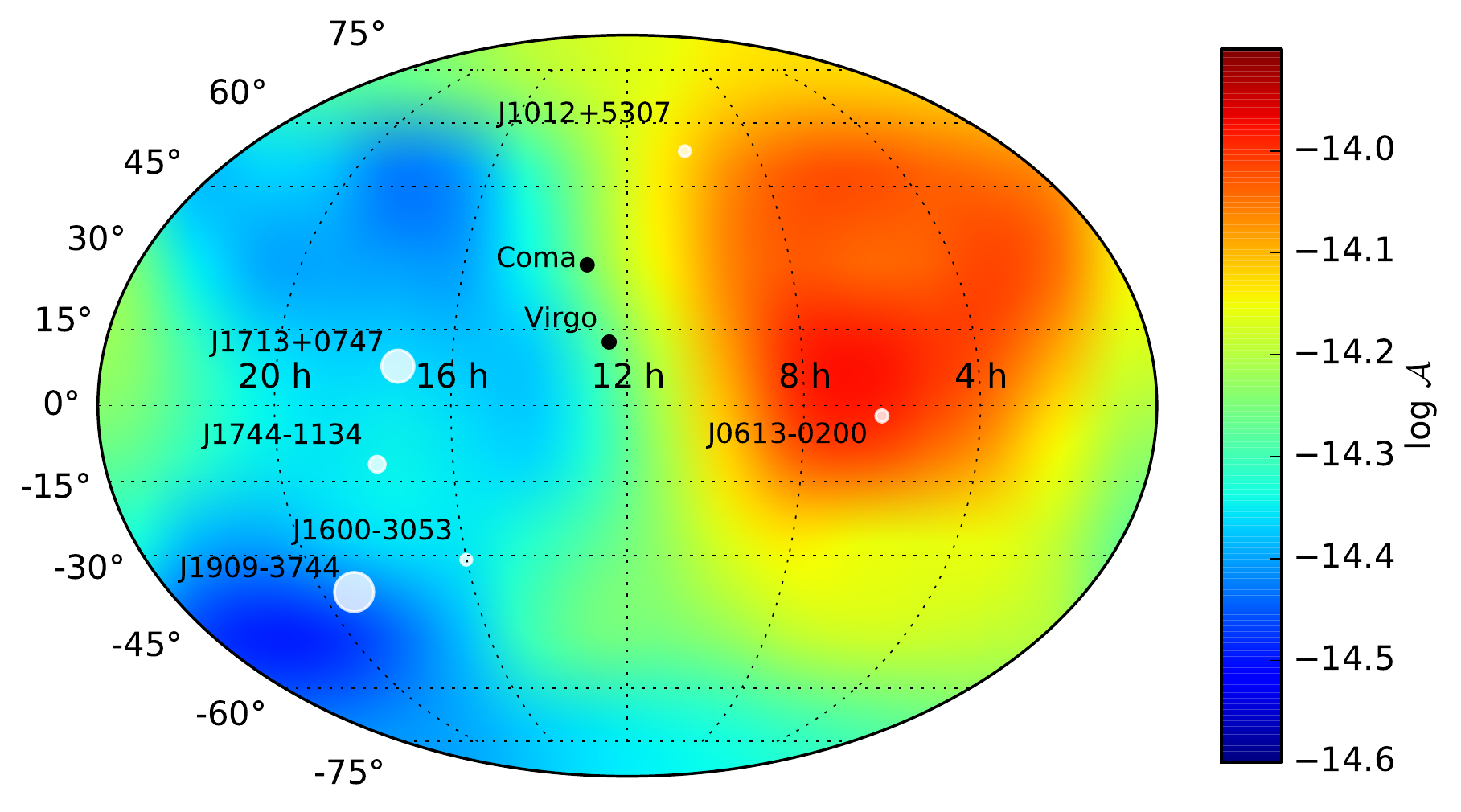}
\caption{ Sensitivity sky map at $f=6.3$ nHz computed with the $\mF_e$ computed with 500 injections in 48 directions in the sky ("cells"). The color scale corresponds to $\log_{10}$ of the 95\% upper limit on the strain amplitude ${\mathcal A}$. The white points indicate the positions of the 6 best pulsars with sizes corresponding to their contribution to the S/N. Black dots indicate the location of the Virgo and the Coma clusters.}
\label{fig:skymapFe}
\end{figure}

In the frequentist framework, this is straightforward to do in the
context of the $\mF_e$-statistic, since it is sensitive to sky
position (unlike $\mF_p$). As already mentioned, the upper limit is
evaluated through massive signal injections according to the procedure
outlined in section \ref{sec:subsubFe}. The difference is that now we
have to divide the sky into ``cells'' and inject 500 sources at each cell
location. This is much more computationally expensive than the
evaluation of the sky-averaged limit; we therefore generate the sky
map at 6.3nHz only, corresponding to our best sky-averaged limit. This
is shown in figure ~\ref{fig:skymapFe}. As expected we are more sensitive 
in the region of the best pulsars.

\begin{figure}
\includegraphics[width=\columnwidth]{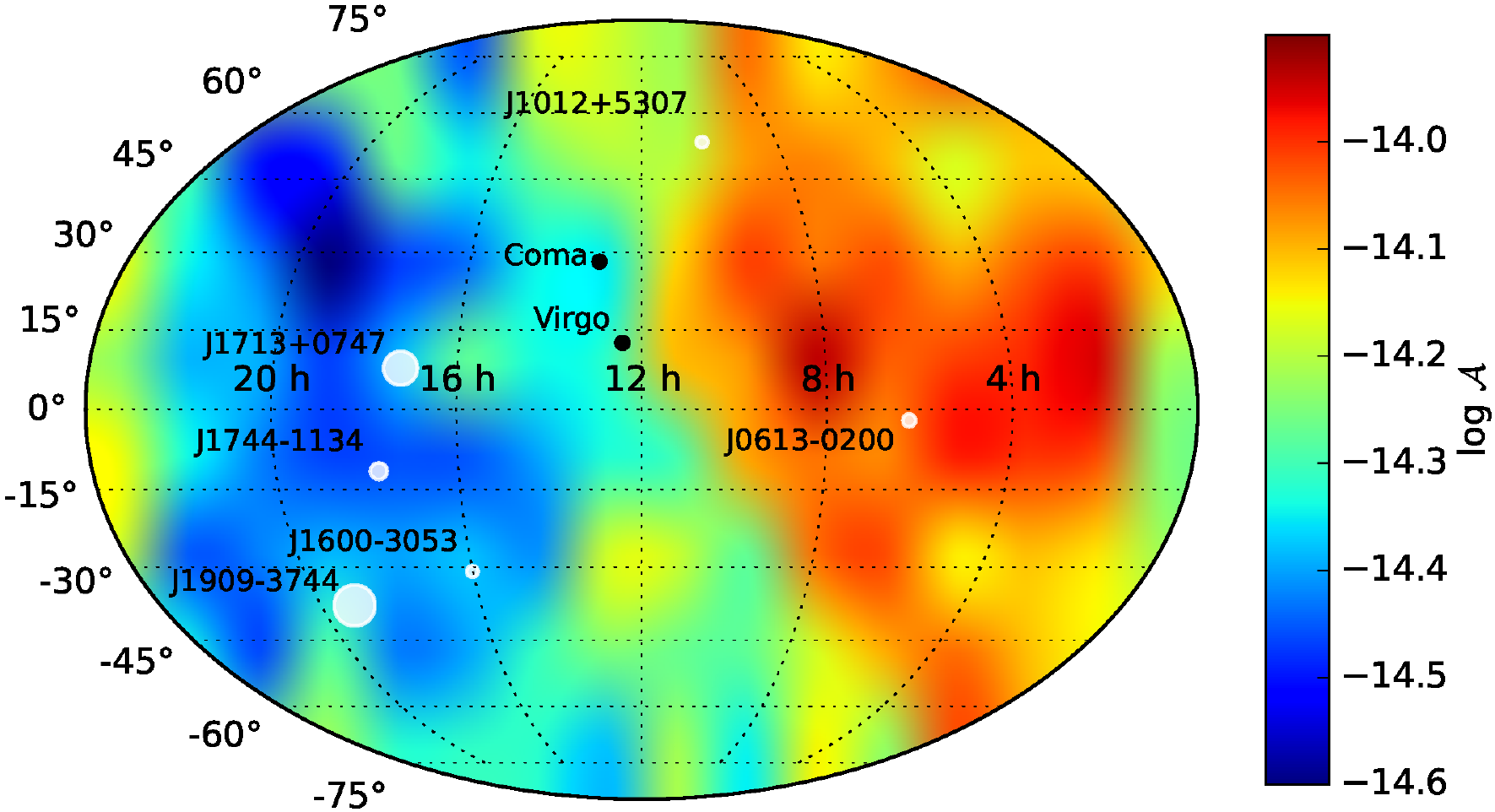}
\caption{Sensitivity sky map at $f=7$ nHz computed with the phase-marginalised Bayesian technique for a ``non-evolving source''. The color scale and points are the same as for figure~\ref{fig:skymapFe}.}
\label{fig:skymap}
\end{figure}

We can also produce targeted upper-limits as a function of
sky location by means of Bayesian techniques. The problem here is that
by splitting the posterior samples on a 3-D frequency-sky location
grid, we end up with only a handful of points per cell, which are not
enough to derive a reliable 95\% upper limit. To mitigate this issue,
we divided the sky into 4$\times$2 patches on the $\phi$ and $\theta$
coordinates, respectively. A dedicated Bayesian analysis (fixed noise
with marginalization of pulsar phases) on each patch yielded enough
samples to sub-divide the region into a further 4$\times$4 sectors, for
a total of 16$\times$8$=128$ resolution elements across the whole
sky. Figure \ref{fig:skymap} illustrates the sky map obtained in this
way at a GW frequency of $7$ nHz (a movie showing the evolution of the
sky map across the relevant frequency range is available at
 http://www.epta.eu.org/aom.html).

The qualitative agreement between the two maps is quite good.
In both of them,
the best pulsars are shown as white dots, with size proportional
to their contribution to the square of the S/N.  As expected, the most
constraining
(i.e. lowest) upper-limits on the strain of a putative CGW lie around
the location of the best pulsars in the array, and the sky maps shows a
clear dipolar pattern. The closest galaxy clusters in the Universe,
i.e. Virgo and Coma, are located at the transition between the two
regions of the dipolar pattern, in an area of ``average sensitivity''.

\section{Astrophysical interpretation}
\label{sec:astro}
The upper limits on CGWs from SMBHBs presented in this paper are currently the most stringent in the literature. We turn now to investigate their impact on the astrophysics of SMBHBs.

\subsection{Horizon distance}
\begin{figure}
\includegraphics[width=\columnwidth]{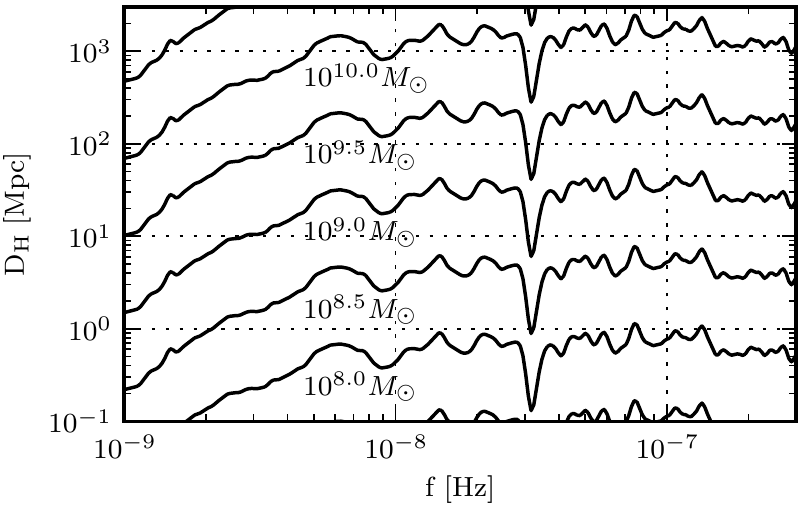}
\caption{Horizon distance as a function of GW frequency for selected values of ${\cal M}_c$, based on the $\mF_p$-{\it ML} upper limit.}
\label{fig:distance_fp}
\end{figure}

Each of the $95\%$ upper limits on ${\mathcal A}$ derived in the previous section can be easily converted into a horizon distance for CGW detection as a function of mass and frequency using equation (\ref{Eq:amplitude}). If ${\mathcal A}_{95\%}(f)$ is the strain upper limit as a function of frequency obtained with a specific method, then
\be
D_H(f,{\cal M}_c) = 2\frac{\mathcal{M}_c^{5/3}}{{\mathcal A}_{95\%}(f)} (\pi f)^{2/3}.
\en
In a frequentist sense, this has to be interpreted as the distance at which, on average, a source of mass $\mathcal{M}_c$ emitting at frequency $f$ located anywhere on the sky would result in a value of the detection statistics {\it higher than what we measure in the data} with $95\%$ probability, if it was there. As an example, results for the $\mF_p$-{\it ML} statistic are presented in figure ~\ref{fig:distance_fp}. An interesting feature of the plot is that, for a given $\mathcal{M}_c$, $D_H$ is essentially constant (slowly declining) for $f>5\times10^{-9}$Hz. This is because of the cancellation effect between the rising CGW amplitude with frequency, ${\cal A}\propto f^{-2/3}$ and the PTA sensitivity, which degrades almost linearly with $f$ (see figures \ref{fig:upperlimit} and \ref{fig:upperlimit_all}). In this frequency range, and with the current sensitivity, we can exclude the presence of a SMBHB with $\mathcal{M}_c>10^9\msun$ out to a distance of about 25Mpc, i.e. well beyond the distance to the Virgo cluster, and with $\mathcal{M}_c>3\times10^9\msun$ out to a distance of about 200Mpc, i.e. twice the distance to the Coma cluster. Note that Virgo and Coma themselves are located in a region of ``average sensitivity'' in our sky sensitivity map (see figure \ref{fig:skymap}), meaning that we can rule out the presence of SMBHBs (with the characteristics described above) in these specific clusters. We remind the reader that these numbers are for SMBHBs with a given {\it redshifted} chirp mass; because of the $1+z$ factor, horizon distances evaluated at the same values of the {\it intrinsic} SMBHB mass will be slightly larger. 

A number of potentially interesting sources have been proposed in the
literature. Among these are the two blazars OJ287
\citep{2008Natur.452..851V} and PG 1302$-$102
\citep{2015arXiv150101375G}. Both objects are located at $z\approx
0.3$ corresponding to $D_L\approx 1.5$Gpc. At such distances, we cannot
rule out any system below $\mathcal{M}_c =10^{10}\msun$, which far
exceeds the plausible range of chirp masses inferred for these two
objects. Should their binary nature be confirmed, these two systems
would likely be among the thousands of contributors to the stochastic
GW background. At this stage, SMBHBs need to be more massive and/or
nearby to be resolved by a PTA.

\subsection{Probability of detection}
\label{subsec:probdet}

A natural question that arises at this point is: could we expect a
detection of a CGW signal with the current EPTA sensitivity? We now
evaluate the probability of detecting CGWs from an individual SMBHB
with an array like the current EPTA, by using a large set of
observationally-based simulations of the cosmic SMBHB population. Each
simulation represents a particular realisation of the ensemble of
SMBHBs. In a nutshell, galaxy merger rates are obtained using a
selection of galaxy mass functions and close galaxy pair fractions
from the literature; merging galaxies are populated with SMBHs
following empirical black hole-galaxy host relations; finally, each
binary is assumed to emit GWs while inspiralling in a quasi-circular
orbit. Given the broad range of different models taken into account
and their uncertainties, numerous simulations are created in order to
cover all possible configurations consistent with the
observations. The SMBHB populations obtained in this way are
consistent with the results of semi-analytic halo merger trees and
cosmological N-body and hydrodynamical simulations. More details on
the simulations and the models employed to produce them can be found
in \cite{sesana13}.

One can perform signal injections drawing the sources from these models and run all the different detection pipelines detailed above, to assess detection probabilities. However, this is an expensive task, and we do not need such a refined analysis at this stage. We instead simplify the problem following a similar approach as in \cite{RosadoEtAl2015}. For a given realization of the SMBHB population, we group GW sources in frequency bins $\Delta{f}=1/T$, and compute the characteristic strain
\begin{align}
\label{eq:hc}
h_c^2=\frac{\sum_k h_k^2 f_k}{\Delta f},
\end{align}
the sum runs over all binaries falling in the frequency bin. We then identify the loudest source in each bin, and compute its S/N following \cite{2010PhRvD..81j4008S} assuming that the noise is given by the sum of the strains of the GWs produced by all other binaries. In practice, we are assuming that all other sources produce an ``unresolved background'', and we check whether the loudest source ``sticks out'' of it. We assume a detection statistic described by a $\chi^2$ distribution with 4 degrees of freedom, and we consider ``individually resolvable'' only those sources  with S/N surpassing the FAP threshold of $0.1\%$ related to this distribution{\footnote{By assuming a $\chi^2$ distribution with 4 degrees of freedom, we are assuming $\mF_e$ as detection statistic. This is an arbitrary choice dictated by computational convenience only. Results are, however, qualitatively unchanged if a different statistic (e.g. $F_p$) is assumed.}}. 

Let us assume a particular upper limit on the GW strain amplitude, among those presented in figure~\ref{fig:upperlimit_all}, and call it $\text{UL}^i$. At a particular frequency bin $f_j$, we simply estimate the probability of detecting a SMBHB with such sensitivity as the fraction of realisations in which a resolvable binary produces a strain amplitude ${\mathcal A}>{\mathcal A}_{95\%}(f_j)$. We call this detection probability $p(\text{D}|\text{UL}^i,f_j)$. The probability of \textit{not} detecting a binary at that frequency is thus $p(\text{N}|\text{UL}^i,f_j)=1-p(\text{D}|\text{UL}^i,f_j)$. Assuming that the probabilities of different frequency bins are independent, the probability of not detecting a binary in any frequency bin is the product of the individual values $p(\text{N}|\text{UL}^i,f_j)$. Hence,
\begin{equation}
p(\text{D}|\text{UL}^i)=1-\prod_j (1-p(\text{D}|\text{UL}^i,f_j))
\end{equation}
is the probability of detecting a SMBHB at any frequency bin, for the upper limit $\text{UL}^i$.

\begin{figure}
\includegraphics[width=\columnwidth]{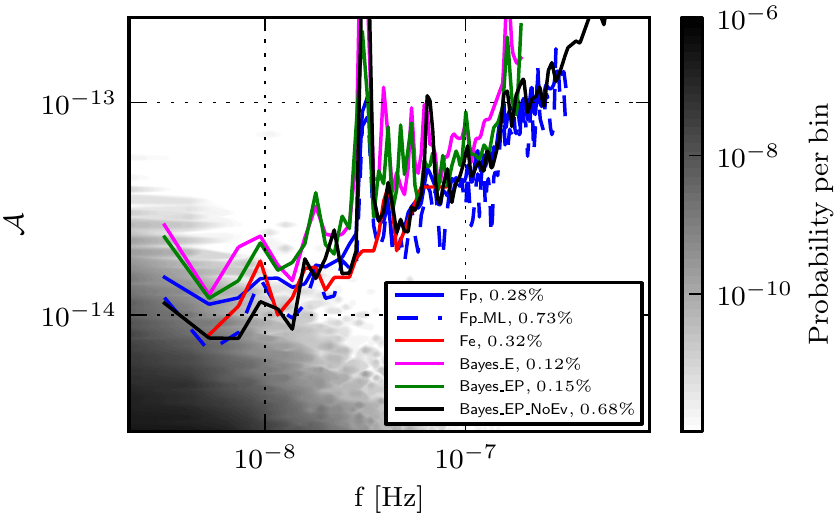}
\caption{GW strain amplitude versus GW observed frequency. The coloured lines represent the different upper limits presented in this work. The shading gives the probability of detecting a SMBHB in a particular interval of strain and frequency. That detection probability increases towards lower frequencies and smaller values of strain (on the lower-left corner). In the legend, the percentage of detection probability is given for each of the upper limits.}
\label{fig:probmodel}
\end{figure}

The detection probability at any frequency obtained for each of the upper limits is given in the legend of figure~\ref{fig:probmodel}.
The maximum detection probability achieved with the EPTA upper limits is below $\sim 1\%$. Therefore, we can safely conclude that a non-detection is consistent with the theoretical expectations.

\subsection{Frequencies of the Earth and the pulsar terms}

In our searches, we distinguished between evolving and non-evolving GW
signals, presenting distinct search methods for each of them. One may
therefore ask, whether one type of signal is more likely than the
other, in order to better focus development efforts on specific
analysis pipelines. We can use the same simulated SMBHB populations
discussed above to answer this question. As shown by
figure~\ref{fig:probmodel}, only a small percentage of them leads to a
detection with the sensitivity of the current EPTA, being therefore
inconsistent with observations. Nevertheless, it is still meaningful
to study the outcome of those realizations, since they would resemble
the true ensemble of SMBHBs in the fortunate case of a detection in
the near future.

\begin{figure}
\includegraphics[width=\columnwidth]{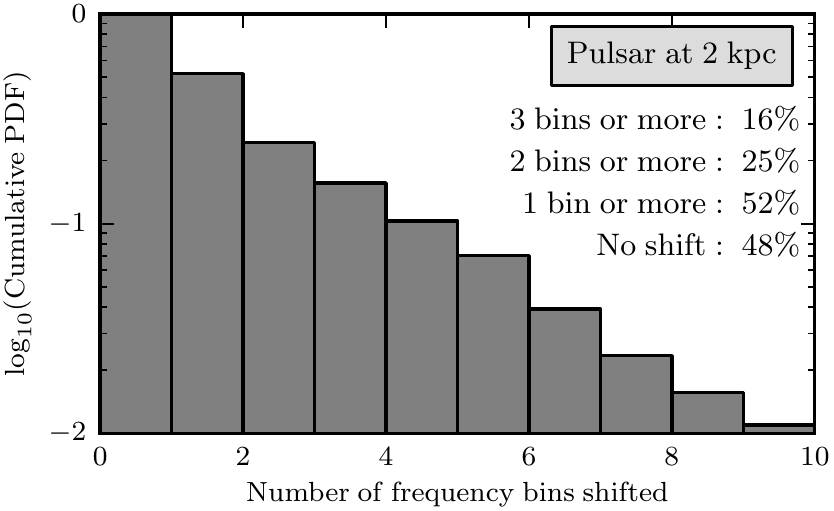}
\includegraphics[width=\columnwidth]{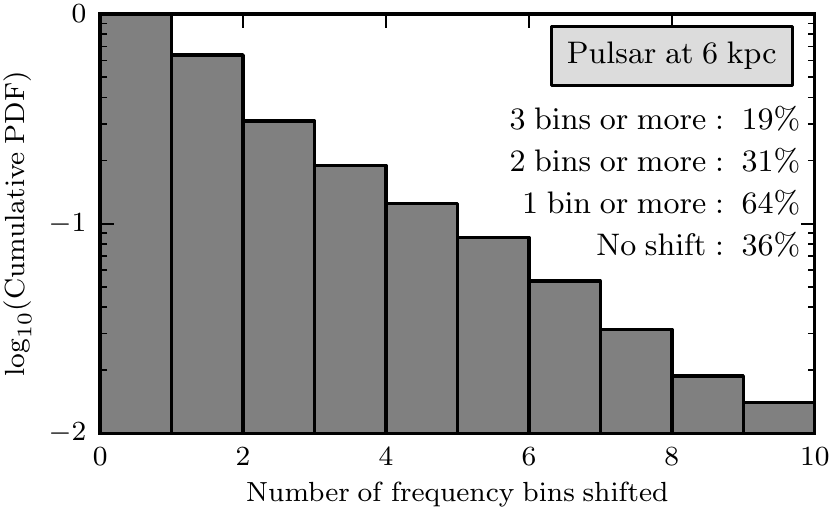}
\caption{Cumulative PDF of the frequency shift between pulsar and Earth terms, in units of the adopted frequency bin $\Delta{f}=$ 15yr$^{-1}$. The upper and the lower panels assume a pulsar distance of 2kpc (average distance of the EPTA pulsars) and 6kpc (maximum distance), respectively.}
\label{fig:fshift}
\end{figure}

For each of the observable SMBHBs in those realizations, we calculate, according to equation (\ref{Eq:freqev}), the frequency evolution of the emitted GW after a time lapse of 4kpc$/c$, which would be the maximum time difference between pulsar and Earth terms for a pulsar located at 2kpc (which is approximately the mean distance to the EPTA pulsars). Frequency shifts are defined as the difference between the GW frequency before and after that time lapse in units of the frequency resolution bin of the array (assumed to be 15yr$^{-1}$ here). Their distribution is shown on the upper plot of figure~\ref{fig:fshift}. The lower plot is analogous, but assuming a pulsar located at 6kpc (which corresponds to approximately the largest distance to a pulsar in the EPTA). 
Should EPTA detect an individual source in the near future, it could either be evolving or non-evolving with nearly equal probability. Even considering the shifts produced in the furthest pulsar only ($\le 6$kpc), there is still a 36\% probability that the source would be non-evolving. Decreasing the PTA sensitivity floor would make it sensitive to lower mass binaries (which evolve faster), but would also improve the chance of detection at higher frequency, where evolution is more likely. Likewise, extending the observation time will allow to see binaries at lower frequency, where evolution is less likely; it will, however, also shrink the size of the frequency bin ($\Delta{f}=1/T$), making it easier for a source to sweep through different resolution elements. Detection strategy development for both classes of sources is therefore warranted\footnote{A more detailed study of the expected properties of the first detectable SMBHBs can be found in \cite{RosadoEtAl2015}.}. 

\section{Conclusions}
\label{sec:conclusions}
   
In this paper, we searched for continuous gravitational wave signals
in the latest release of the European Pulsar Timing Array dataset. We
adopted both frequentist and Bayesian techniques, searching for both
frequency-evolving and strictly monochromatic signals. In most of the
cases, we fixed the value of the noise parameters in each pulsar to
the maximum likelihood estimated in a separate single-pulsar
analysis. This choice was primarily dictated by computational
feasibility, but is certainly non optimal, since pulsar noise and GW
signals might be degenerate and one should include both simultaneously
in the search. To validate our results we therefore also performed a
frequentist analysis, sampling from the posterior distributions of the
noise parameters returned by the SPA (simply labeled $\mathcal{F}_e$), and a
full Bayesian search over both the noise and the signal (labeled {\it
  Bayes\_EP\_NoEv\_noise}). Because of the high dimensionality of the
search parameter space, the latter has been conducted on a
restricted dataset including only the 6 best pulsars in the array.

None of the analysis yielded any evidence of the presence of a signal, and only upper limits on the amplitude ${\cal A}$ of a putative CGW could be placed. The 
 excellent quality and length of the dataset allowed us to place limits comparable to those of \cite{2014MNRAS.444.3709Z} at $f>10$nHz and a factor of two better at $f<10$nHz, yelding the overall most stringent constrains to date. All the employed methods yield 95\% upper limits on ${\cal A}$ (${\cal A}_{95\%}$) consistent within a factor of two across the whole 2nHz-400nHz frequency range. Our best sensitivity is in the 5nHz-7nHz interval, where we find $6\times10^{-15}<{\cal A}_{95\%}<1.5\times10^{-14}$, depending on the adopted method. The most robust analysis ({\it Bayes\_EP\_NoEv\_noise}) results in ${\cal A}_{95\%}=9\times10^{-15}$ at 6nHz. Limits on the strain amplitude can be converted to horizon distances as a function of source mass and frequency. We exclude the existence of  SMBHBs with  separation $<0.01$pc and $\mathcal{M}_c>10^9\msun$ out to a distance of about 25Mpc (well beyond Virgo),  and with $\mathcal{M}_c>10^{9.5}\msun$ out to a distance of about 200Mpc (twice the distance to Coma). In recent years, several ``overmassive'' black holes have been found in the local Universe, with measured masses in excess of $10^{10}\msun$. Our analysis excludes that any such system lives in a compact binary within a distance of about 1Gpc ($z\approx0.2$). Finally, we compared our limits to the predictions of state of the art models of the cosmic population of SMBHBs. We found a detection probability of $\lesssim1\%$ at current sensitivity, consistent with the null result of our searches.

The present analysis has also highlighted a few interesting technical
issues related to the search methods and to the nature of the
dataset. Despite not being robust for detection purposes, as pointed
out by \cite{2014ApJ...794..141A}, fixed noise analysis upper limits
are consistent within 50\% of those obtained by searches over the
full parameter space (i.e., including signal and noise
simultaneously). Therefore, so long as the data do not support the
presence of a signal, a computationally cheap analysis of this type
can be carried out over an extensive dataset of numerous pulsars,
possibly yielding more interesting astrophysical constraints on the
low-redshift SMBHB population in the near future. Eventually,
simultaneous searches over the signal and noise parameters will be
required for a confident detection claim. However, those are extremely
expensive, and novel techniques capable of efficiently handling
parameter spaces of 100+ dimensions must be developed. The reason why
the results of the full search on the restricted dataset of 6 pulsars
is consistent with those provided by fixed noise analysis on the full
set of 41 pulsars, is that the current EPTA array is heavily dominated
by a handful of ultra-stable MSPs. In particular, PSRs J1909$-$3744
and J1713+0747 combined account for 80\% of the EPTA sensitivity to
CGWs. As a result, the EPTA dataset sensitivity has a strongly dipolar
pattern across the sky, varying by almost a factor of four over the
celestial sphere. The discovery of new ultra-stable MSPs will
therefore be crucial to provide a better sky coverage, ensuring that
no ``blind spots'' are left, and thus enhancing the probability of
detecting CGWs in the coming decade.

\section*{Acknowledgments}
Part of this work is based on observations with the 100-m telescope of the Max-Planck-Institut f{\"u}r Radioastronomie (MPIfR) at Effelsberg. The Nan{\c c}ay radio Observatory is operated by the Paris Observatory, associated to the French Centre National de la Recherche Scientifique (CNRS). We acknowledge financial support from 'Programme National de Cosmologie and Galaxies' (PNCG) of CNRS/INSU, France.   Pulsar research at the Jodrell Bank Centre for Astrophysics and the observations using the Lovell Telescope is supported by a consolidated grant from the STFC in the UK. The Westerbork Synthesis Radio Telescope is operated by the Netherlands Institute for Radio Astronomy (ASTRON) with support from The Netherlands Foundation for Scientific Research NWO. 

This research was performed using several supercomputers~: the CCIN2P3 computer cluster of the CNRS-IN2P3 (Lyon-France), the ARAGO computer cluster of the François Arago Centre (Paris-France), the Darwin Supercomputer of the University of Cambridge High Performance Computing Service (http://www.hpc.cam.ac.uk/), provided by Dell Inc. using Strategic Research Infrastructure Funding from the Higher Education Funding Council for England and funding from the Science and Technology Facilities Council, and the Vulcan cluster of MPIfG- AEI (Golm-Germany).
The authors acknowledge the support of VirtualData from LABEX P2IO for providing computing resources through its StratusLab cloud.
\cR{This work was supported in part by the National Science Foundation under Grant No. PHYS-1066293 and the hospitality of the Aspen Center for Physics.}

LL was supported by a Junior Research Fellowship at Trinity Hall College, Cambridge University.  ST was supported by appointment to the NASA Postdoctoral Program at the Jet Propulsion Laboratory, administered by Oak Ridge Associated Universities through a contract with NASA. CMFM was supported by a Marie Curie International Outgoing Fellowship within the 7th European Community Framework Programme.  AS and JG are supported by the Royal Society.  SAS acknowledges funding from an NWO Vidi fellowship (PI JWTH).  RNC acknowledges the support of the International Max Planck Research School Bonn/Cologne and the Bonn-Cologne Graduate School.  KJL is supported by the National Natural Science Foundation of China (Grant No.11373011).  RvH is supported by NASA Einstein Fellowship grant PF3-140116.  JWTH acknowledges funding from an NWO Vidi fellowship and ERC Starting Grant `DRAGNET' (337062).   PL acknowledges the support of the International Max Planck Research School Bonn/Cologne. SO is supported by the Alexander von Humboldt Foundation.

\bibliographystyle{mn2e}
\bibliography{references}

\bsp

\label{lastpage}

\end{document}